\documentclass[sn-mathphys,Numbered]{sn-jnl}

\usepackage{framed,multirow}
\usepackage{graphicx}%
\usepackage{multirow}%
\usepackage{amsmath,amssymb,amsfonts}%
\usepackage{amsthm}%
\usepackage{mathrsfs}%
\usepackage[title]{appendix}%
\usepackage{xcolor}%
\usepackage{textcomp}%
\usepackage{manyfoot}%
\usepackage{booktabs}%
\usepackage{algorithm}%
\usepackage{algorithmicx}%
\usepackage{algpseudocode}%
\usepackage{listings}%

\usepackage{booktabs}
\usepackage{array}
\usepackage{pifont}
\usepackage{tabularx}
\usepackage{bbding}
\newcommand{\cmark}{\ding{51}}
\newcommand{\xmark}{\ding{55}}
\theoremstyle{thmstyleone}%
\theoremstyle{thmstyletwo}%

\theoremstyle{thmstylethree}%

\begin{document}

\title[Article Title]{LoLiSRFlow: Joint Single Image Low-light Enhancement and Super-resolution via Cross-scale Transformer-based Conditional Flow}


\author[1]{\fnm{Ziyu} \sur{Yue}}\email{11901015@mail.dlut.edu.cn}

\author[2]{\fnm{Jiaxin} \sur{Gao}}\email{jiaxinn.gao@outlook.com}
\author[2]{\fnm{Sihan} \sur{Xie}}\email{XSH2018@mail.dlut.edu.cn}

\author[3]{\fnm{Yang} \sur{Liu}}\email{liu@nudt.edu.cn}

\author*[1]{\fnm{Zhixun} \sur{Su}}\email{zxsu@dlut.edu.cn}

\affil*[1]{\orgdiv{School of Mathematical Science}, \orgname{Dalian University of Technology}, 
		 \orgaddress{
		\city{Dalian}, \postcode{116024}, \state{LiaoNing}, \country{China}}}

\affil[2]{\orgdiv{School of Software Technology}, \orgname{Dalian University of Technology},
	 \orgaddress{
		\city{Dalian}, \postcode{116024}, \state{LiaoNing}, \country{China}}}

\affil[3]{\orgdiv{Department of Mathematics}, \orgname{National University of Defense Technology},
			 \orgaddress{
		\city{Changsha}, \postcode{410073}, \state{HuNan}, \country{China}}}  


\abstract{The visibility of real-world images is often limited by both low-light and low-resolution, however, these issues are only addressed in the literature through Low-Light Enhancement (LLE) and Super- Resolution (SR) methods. Admittedly, a simple cascade of these approaches cannot work harmoniously to cope well with the highly ill-posed problem for simultaneously enhancing visibility and resolution. In this paper, we propose a normalizing flow network, dubbed LoLiSRFLow, specifically designed to consider the degradation mechanism inherent in joint LLE and SR. To break the bonds of the one-to-many mapping for low-light low-resolution images to normal-light high-resolution images, LoLiSRFLow directly learns the conditional probability distribution over a variety of feasible solutions for high-resolution well-exposed images. Specifically, a multi-resolution parallel transformer acts as a conditional encoder that extracts the Retinex-induced resolution-and-illumination invariant map as the previous one. And the invertible network maps the distribution of usually exposed high-resolution images to a latent distribution. The backward inference is equivalent to introducing an additional constrained loss for the normal training route, thus enabling the manifold of the natural exposure of the high-resolution image to be immaculately depicted. We also propose a synthetic dataset modeling the realistic low-light low-resolution degradation, named DFSR-LLE, containing 7100 low-resolution dark-light/high-resolution normal sharp pairs. Quantitative and qualitative experimental results demonstrate the effectiveness of our method on both the proposed synthetic and real datasets.}

\keywords{Low-light Vision, Image Enhancement, Low-light Image Super-resolution, Normalization Flow}

\maketitle

\section{Introduction}\label{sec1}
High-quality images with richly detailed information are essential for visual learning tasks~(\cite{liu2018deep,zhang2019two,smeulders2013visual}). However, real-world images captured in low-light environments usually suffer multiple degradations, such as low-resolution and poor visibility.
Prior methods address the two image processing tasks independently, i.e., low-light enhancement~(\cite{liu2021retinex,guo2020zero,guo2016lime}) and image super-resolution~(\cite{lugmayr2020srflow,cai2019toward,wang2018esrgan}). These methods guide the degradation learning process according to each specific physical model and independent assumptions in their specific tasks. 
As a result, any individual task-specific approach cannot effectively cope with this joint degradation problem. 
As shown in Figure~\ref{fig:figure1}, existing LLE and SR methods applied directly to the joint task often fall into collapse, may either result in noise amplification and destruction of scene detail, or fail to improve illumination in low-light images, leading to various luminance deficits and unexplained artifacts.
Further, we find that a simple combination of two task-specific augmentation methods does not substantially solve this complex degradation problem, but instead causes further noise and artifact amplification. Also, it is experimentally demonstrated that performing LLE first and then SR results in significant amplification of noise and blur, with various noises and artifacts. The reverse order of execution also leads to improper brightness and low color saturation.
We intuitively infer that the main reason for this series of issues may be that the existing state-of-the-art methods are designed for a single degradation factor and insufficiently characterize the essence of the super-resolution problem in low-light scenes.

Recently, researchers have demonstrated the effectiveness of normalizing flow for highly ill-posed problems in vision and learning domains~(\cite{lugmayr2020srflow,winkler2019learning}).  Through constructing a series of cascaded reversible network layers, the normalizing flow can implement the conversion between the expected distribution of the real image and the simple distribution of the latent space. 
However, since classical normalization flow usually does not consider the intrinsic properties of the task, applying these methods to LLE and SR co-processed images may fail to model the properties of the image itself (e.g., color, illuminance, etc.) can cause problems such as color bias, artifacts, and so on.

To address the above issues, in this paper, we aim to train a specialized network to solve the joint task involving two degradation types. Recently, researchers have demonstrated the effectiveness of normalizing flow for highly ill-posed problems in vision and learning domains~(\cite{lugmayr2020srflow,winkler2019learning}).  Through constructing a series of cascaded reversible network layers, the normalizing flow can implement the conversion between the expected distribution of the real image and the simple distribution of the latent space. 
However, since the classical normalizing flow generally does not take into account the intrinsic properties of the task, it may fail to model some global image properties like color saturation, which can undermine the performance when applying these methods for joint LLE and SR image processing problems. 

Rather than learning a deterministic one-to-one mapping between a low-light low-resolution image and its ground truth, we propose a Low-Light Super-Resolution Flow (LoLiSRFlow) network based on the normalizing flow to learn the conditional distribution of a normally exposed high-resolution image. Specially, we introduce a resolution- and illumination-invariant color ratio map as a prior and take it as the mean of the Gaussian distribution, which means that color and texture distributions are consistent for different resolutions and darkness levels. In this way, it can be understood that the conditional encoder aims to learn resolution and illumination invariant intrinsic properties. 
At the same time, a multi-scale transformer is introduced as a conditional encoder to accurately learn resolution- and light-invariant features as well as priors. 
During the training phase, high-resolution normal exposure images are mapped into the the implicit latent space through the inverse of a invertible flow-based model network. In the inference stage, by sampling the color ratio map according to the highest probability value, the optimal reconstruction image can be obtained through the reversible flow. Additionally, to address the dataset scarcity problem, we also propose a new dataset through a low-luminance low-resolution image synthesis pipeline that is designed to approximate the degradation process of real-world data. Extensive experiments demonstrate the superiority of LLSRFlow on proposed synthetic dataset and real-world data.  

The main contributions are summarized as follows:
\begin{itemize}
	\item We analyze the necessities of jointly addressing low-light enhancement and super-resolution tasks, and propose an effective LoLiSRFlow method upon the inherent degradation process of this joint task.
	\item We propose a novel transformer-based network via introducing the conditional normalizing flow in a paralleled multi-resolution manner. To regularize the learning process and image manifold of the proposed network such that a high consistency in both color and content can be well preserved, we introduce a novel resolution- and illumination-invariant color ratio map served as the prior.
	\item  We propose a new dataset named DFSR-LLE dataset for the joint low-light enhancement and super-resolution task, which contains $7100$ pairs of images covering various scenes with real noise model in ISP process.  
\end{itemize}

\begin{figure*}[htbp]
	\centering 
	\begin{tabular}{c}
		\includegraphics[height=6.8cm,width=12.5cm]{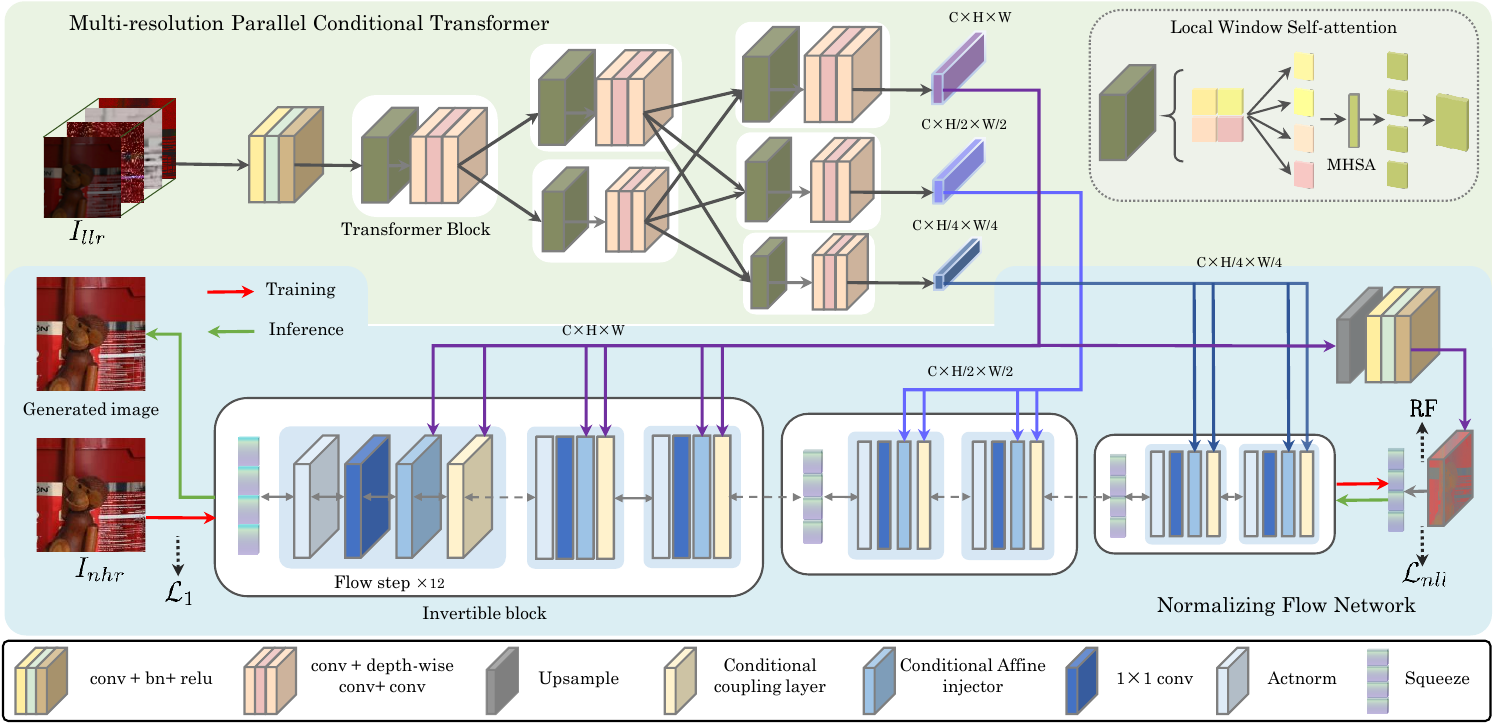}\\
	\end{tabular}%
	\caption{An illustration of the proposed LoLiSRFlow.  Our model consists of a multi-resolution parallel conditional  transformer (MPCT) to extract the cross-scale illumination-invariant reflection map and an normalizing flow based invertible network that learns a distribution of high-resolution normally exposed images conditioned on a low-resolution low-light one. }\label{fig:llsr}
\end{figure*}%

 \section{Related works}
 In recent years, deep learning methods for solving super-resolution and low illumination enhancement tasks have emerged as deep learning has been applied to image restoration.
 This section summarizes recent trends for two types of image processing tasks and the development of normalizing flow architectures. 
 
 \subsection{Image Super-resolution} 
 Currently, many Convolutional Neural Network (CNN) and Generative Adversarial Network (GAN) based methods have achieved surprising results in image super-resolution~(\cite{ledig2017photo,zamir2020learning,wang2018esrgan}). 
 ESRGAN~(\cite{wang2018esrgan}) improves the adversarial and perceptual losses and constructs a generator based on Residual Dense Block (RRDB) and removes the batch normalization operation from the network module. 
 To address the problem of image super-resolution in real scenes, RealESRGAN~(\cite{wang2021real}) further improves the discriminator structure and proposes a high-order degradation model.
 MIRnet~(\cite{zamir2020learning}) has a good effect in the field of image enhancement including SR which combines the attention mechanism to capture spatial and channel context information, and extract complementary features at multiple scales. 
 Compared with the CNN methods GAN methods, the field of SR is also researched on transformer based network. The advent of Swin transformer~(\cite{liu2021swin}) attracts lots of attention. For example, inspired by Swin transformer, Liang et.al~(\cite{liang2021swinir}) constructs SwinIR network as a baseline model for various image restoration tasks. 
 It has the advantages that image content and attention weights based on content interaction can be regarded as spatial variable convolution, and the shift window mechanism in Residual Swin Transformer Block (RSTB) can perform long-distance dependency modeling.
 HAT~\cite{chen2023activating} proposes a hybrid attention transformer. it combines the mechanisms of channel attention and window-based self-attention so as to take advantage of their complementary strengths of being able to utilize global statistics and stronger local fitting capabilities, and to be able to better aggregate cross-window information and enhance detail.
 
 \subsection{Low-light Enhancement}
 Early LLE methods mainly use Retinex theory to correct the illumination of the image and suppress artifacts. Based on the Retinex theory, the low-light image can be decomposed into the incident component and the reflection component which should be obtained to recover the incident as the enhancement result.
 LIME~(\cite{guo2016lime}) estimates the image illumination map based on the bright channel prior and the structure prior, and finally synthesizes the enhanced image according to the Retinex theory.
 Research of RUAS~(\cite{liu2021retinex}) takes neural network structure search (NAS)~(\cite{liu2018darts}) as the core method, introduces prior constraints based on Retinex theory to establish a LLE model. 
 and discovers an efficient deep network structure for low-light prior from a custom compact search space, and by expanding its optimization solution process to build the overall network structure. 
 However, the results obtained are usually unnatural and prone to over-enhancement.
 ZeroDCE~(\cite{guo2020zero}) learns quadratic curves of enhanced images by designing a sophisticated non-reference loss function in the absence of reference images, such as spatial consistency loss, exposure control loss, color constancy loss, illumination smoothness loss. Although it has a good effect on enhancing brightness, it does not consider the problem of noise amplification. 
 URetinex-Net~(\cite{wu2022uretinex}), on the other hand, proposes a deep unfolding network based on the retinex theory, which unfolds the optimization problem into a learnable network to solve the low-light enhancement problem.
 More recently, the more lightweight and robust low-light enhancement method SCI~(\cite{ma2022toward}) has been proposed, using only a single basic module in the inference phase and an unsupervised training strategy to further improve the scene generalization of the model.
 
 \subsection{Normalizing Flow} 
 
 The normalized flow, as a type of deep generative model, learns an invertible transformation of an implicitly encoded distribution to an image domain distribution by means of log-likelihood constraints.
 The generative process can be specified by an invertible architecture from a latent variable. Since the learned transformations are reversible, there is no loss of information when the two distributions are transformed into each other.  
 A series of pioneering works~(\cite{dinh2014nice,kingma2018glow,jacobsen2018revnet,ardizzone2019guided}) have been devoted to the elaborate design of the layers of the network in order to make the network invertible, as well as to make the determinant of the Jacobi matrix computationally tractable. 
 Some recent works exploit conditional normalizing flows to improve the expressiveness~(\cite{trippe2018conditional,winkler2019learning,sun2019dual,pumarola2020c,atanov2019semi,zhao2021invertible}).  
 In the super-resolution tasks, Wolf et al.~(\cite{wolf2021deflow}) uses the distribution of high-frequency information from high-resolution images as the distribution in the implicit coding space and jointly learns the distributional transform from low-resolution images to high-resolution images. SRFlow~(\cite{lugmayr2020srflow}) introduces low-resolution images as conditional information into the normalized flow architecture, which learns the distribution of realistic high-resolution images. 
 In this paper, we integrate normalizing flow as our basic architecture to improve the performance of joint LLE and SR models. 
 
 \section{Methodology}
 In this section, we propose a transformer-based conditional normalizing flow network to directly learn a conditional distribution of normal-light high-resolution images conditioned on  low-light low-resolution input. 
 
 \subsection{Preliminary} 
 \noindent\textbf
 {Problem definition.}
 The goal of the joint LLE and SR task is to reconstruct a normal-light, high-resolution image $\mathbf{y}$ from its low-light, low-resolution counterpart $\mathbf{x}.$ 
 Unlike most existing deterministic mapping methods from $\mathbf{x}$-to-$\mathbf{y}$, we aim to learn a conditional probability distribution $P_{\mathbf{y}|\mathbf{x}}(\mathbf{y}|\mathbf{x};\theta)$ of a normally exposed high resolution corresponding to a low-light, low-resolution input. 
 The core idea is to use information non-destructive network modeling of reversible flow models $F_{\theta}$ to represent the conditional probability density $P_{\mathbf{y}|\mathbf{x}}.$
 In the training stage, the invertible normalizing flow network takes a low-resolution low-light image $\mathbf{x}$ as input and outputs the latent variable $z$ with the same dimension of $ \mathbf{y} $, i.e., $z=F(\mathbf{y}|\mathbf{x};\theta_{F}).$ 
 By using the change of variable theorem, the conditional probability density distribution of a normally exposed image can be expressed as:
 \begin{equation}
 	P_{\mathbf{y}|\mathbf{x}}(\mathbf{y}|\mathbf{x};\theta)=P_z(F_{\theta}(\mathbf{y},\mathbf{x}))\left| det (\mathcal{J}_ F) \right|,
 \end{equation}
 where $\mathcal{J}_ F$ denotes the Jacobian, i.e.,  $\frac{\partial F_{\theta}(\mathbf{y},\mathbf{x})}{\partial \mathbf{y}}$.
 
 Unlike general generative adversarial networks that minimize the divergence function, the flow model directly optimizes the probability density function. That is, we can train the flow architecture via maximizing $P_{\mathbf{y}|\mathbf{x}}(\mathbf{y}|\mathbf{x};\theta)$, which is equivalent to minimizing the negative log-likelihood loss $L_{nll}$:
 \begin{equation}
 	\begin{aligned}	
 		L_{nll}(\mathbf{y},\mathbf{x};\theta)&=-logP_{\mathbf{y}|\mathbf{x}}(\mathbf{y}|\mathbf{x};\theta) \\ &=-logP_z(F_{\theta}(\mathbf{y},\mathbf{x}))-log\left| det (J_ F) \right|.\\
 	\end{aligned}
 \end{equation}
 Here the invertible network $F_{\theta}$ consists of a series of sequentially coupled invertible layers, i.e., 
 $F_{\theta}:=F_{\theta}^{N-1} \circ \cdots F_{\theta}^{1}\circ  F_{\theta}^{0}$, where $\circ$ denotes the compound of functions. 
 Let $g^{n}$ represents the output of reversible layer $F_{\theta}^{n}$. Each reversible layer of the flow model in the training process, accepts the information from the previous layer and conditional encoder. Thus the calculation process of each reversible layer is expressed as $g^{n+1}=F_{\theta}^{n}(g^{n},E^{n}_{\theta})$\footnote{Here $E^{n}$ keeps the same dimension as $g^{n}$ in both training and inference stages.}, where $g^{0}=\mathbf{y}$, $g^{N}=z,$ $E$ is the conditional encoder.  Thus, the negative log-likelihood loss function can be expressed as:
 \begin{equation}
 	\begin{aligned}
 		L_{nll}(\mathbf{y},\mathbf{x};\theta)=&-logP_z(F_{\theta}(\mathbf{y},\mathbf{x}))\\
 		&-\sum_{n=0}^{N-1}log\left|det \frac{\partial F_{\theta}^{n}(g^{n},E^{n}_{\theta})}{\partial g^{n}} \right|.\label{con:nnlloss}
 	\end{aligned}
 \end{equation}
 From the above equation, the Flow model computes the Jacobian log-determinant of each invertible layer throughout the training phase and only needs to optimize the negative log-likelihood to learn the conditional distribution.  
 
\begin{figure}[htb]
 	\begin{center}
 		\begin{tabular}{c}
 			\includegraphics[height=3.26cm,width=8.25cm,trim=10 0 10 0,clip]{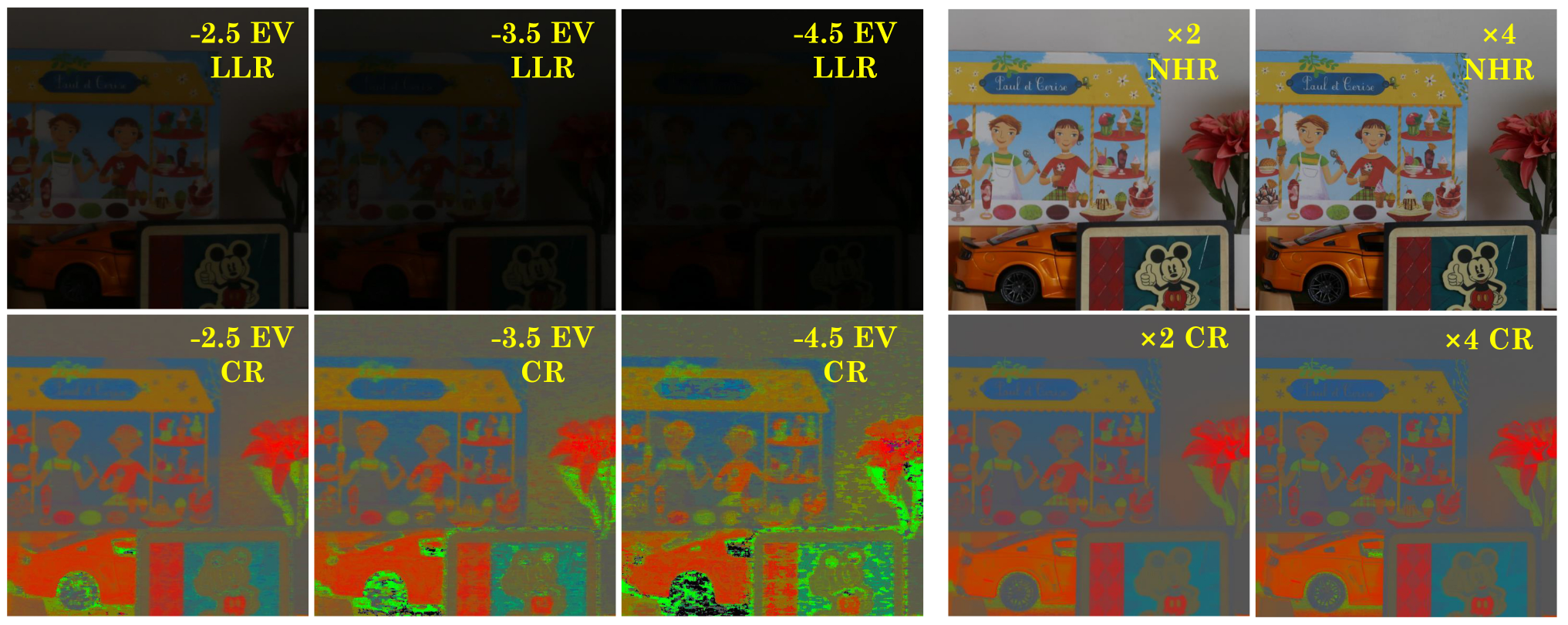}\\
 		\end{tabular}
 	\end{center} 
 	\caption{Visual results of color ratio map $CR(I)$ on the low-light low-resolution image with three darkness levels (i.e., $-2.5$ EV, $-3.5$ EV and $-4.5$ EV),  and normal-light high-resolution image with two scales ($\times2$, $\times4$).}
 	\label{fig:CR}
 \end{figure}

 \noindent\textbf{Resolution- and illumination-invariant map.} Given a image $I$, we propose to calculate its Retinex-induced color ratio map $CR(I)$, i.e., 
 $CR(I)=\frac{I}{Sum\left( I\right)}$ \footnote{Note $Sum$ denotes the summation function that calculates each pixel among three channels of $I$.}.
 It can be regarded as an a prior information of the image that does not change with the degree of illumination and the resolution of the image.
 In the training phase, we randomly sample a data pair $\left\lbrace \mathbf{x},\mathbf{y}\right\rbrace $ from the real dataset~(\cite{aakerberg2021rellisur}) with three different levels of darkness due to different exposure times. 
 Also, we calculate the color ratio map for the low-light low-resolution image $ \mathbf{x}$ with three level of exposure, and corresponding normal-light high-resolution image $\mathbf{y}$ among two scales ($\times2,\times4$), respectively. As shown in Figure~\ref{fig:CR}, the results show that the color ratio map $CR(I)$ remains consistent across image scales and exposures, which can be regarded as an inherent property of invariant resolution and illumination (corresponding to the reflection map in the Retinex model). 
 
 \subsection{LoLiSRFlow}
 Rather than previous CNN-based conditional normalizing flow, we construct a multi-resolution parallel transformer-based conditional flow model for joint LLE and SR, called LoLiSRFlow. As is illustrated in Figure~\ref{fig:llsr}, LoLiSRFlow includes two main modules: a multi-resolution parallel transformer-based conditional encoder $E$ takes a low-light low resolution image $\mathbf{x}$ as input and outputs resolution- and illumination-invariant color ratio map as well as conditional feature maps of different scales, and an invertible normalizing flow network $F$ that maps a normal-light high-resolution image $\mathbf{y}$ to a latent encoding $z$.  
 
 \noindent\textbf{Multi-resolution parallel conditional encoder.} Considering the intrinsic degradation mechanism of the joint task, the conditional encoder learns a one-to-one mapping to depict the task inherent property, that is, to learn a mapping from a low-light low-resolution input $\mathbf{x}$ to the color ratio map. 
 Experiments show that the encoders constructed by traditional residual convolutional network architectures (e.g., Residual in Residual Dense Block, RRDB~(\cite{zhang2018residual})) often lose high-frequency information and produce excessive artifacts. Thus, inspired by~(\cite{yuan2021hrformer}), we design a multi-scale transformer-based encoder to fuse multi-resolution feature information, which mainly consists of a local-window self-attention module, feed-forward network with depth-wise convolution and feature blending module in Figure~\ref{fig:llsr}.  
 
 \noindent\textbf{Invertible normalizing flow network.}As a condition generator, the normalizing flow network consists of $3$ scale invertible blocks, and each block contains $12$ flow steps that are shared with the same $4$ flow layers. In each flow step, there are also two extra operations, i.e., squeeze and split. 
 The squeezing operation reorders the features input to the next level of the reversible flow block at different scales, decreasing their resolution and increasing the number of feature channels.
 The split operation, on the other hand, disrupts the channel order of the features and divides them into two parts by reversible convolution before the features enter each flow module, allowing the feature information to be more fully fused in each flow block.
 Similar to the generic architecture for flow modeling~(\cite{lugmayr2020srflow}), the each reversible flow block we use consists of four parts, which are the actnorm layer, the $1 \times 1$ invertible convolution, the conditional coupling layer, and the affine injector.

 \noindent\textbf{Loss Function.} To train our proposed LLSRFlow, we utilize the negative log-likelihood loss $L_{nll}$ and $L_1$ loss simultaneously. 
 In order to suppress the structural distortion of the edge and texture information, we use the pixel-wise reconstruction loss $L_1$ to measure the difference between the latent image $z$ and the normal-light high-resolution image $\mathbf{y}$, given by
 \begin{equation}
 	L_1 = ||F_{\theta}^{-1}(z/\mathbf{x})-\mathbf{y}||_1.
 \end{equation}
 The negative log-likelihood loss is defined in Eq~\ref{con:nnlloss}, where the probability distribution function $P_{z}(z)$ is re-represented as:
 %
 \begin{equation}
 	P_{z}(z)=RF\left(\frac{1}{\sqrt{2\pi}}e^{-\frac{\left(z-E_{\theta}(\mathbf{x})\right)^2}{2}}, \frac{1}{\sqrt{2\pi}}e^{-\frac{\left(z-CR(\mathbf{y})\right)^2}{2}}\right).
 \end{equation}
 Note that $ RF$ is defined as a random function taking values among two dimensions, and in this paper the two coordinate values are chosen with a probability of $ 1:4 $.
 The overall loss is a trade-off of two loss terms:
 $ 
 L_{total}=L_{nll} +\gamma L_1, 
 $ 
 where $\gamma$ denotes the weighting factor set as $1.5$. 
 
 \subsection{The proposed DFSR-LLE Dataset}
 In order to address the data scarcity problem for the joint task, we build a new dataset named DFSR-LLE based on DIV2K ~(\cite{Agustsson_2017_CVPR_Workshops}) and Flickr2K ~(\cite{timofte2017ntire}). The DFSR-LLE dataset contains 7100 pairs of low-resolution, low-light, and high-resolution, normal light data, and is divided into 2x and 4x resolution scales. Specifically, the 2x and 4x data in the proposed dataset each have 3550 pairs, and the data used for training and testing are 3350 and 200 pairs, respectively. Based on real camera imaging, we first adjust the darkness on the original high-resolution, normally light image. A bicubic downsampling operation is then performed on the darkened image, resulting in a low-resolution, low-light image. Finally, to attach near-real camera noise to the image, we add Shot and Read Noise to the image, resulting in a final low-resolution, low-light image. In the following we will describe the specific process of obtaining synthetic data.
 \vspace{0.05cm}
 
 \noindent\textbf{Adjusting darkness.} In order to adjust the brightness consistently of high-resolution, normal lighting data, we define 
 $I^{(c)}_{out}=\beta\times(\alpha\times I^{(c)}_{in})^{\gamma}, c\in {R, G, B},$
 where $\alpha$ and $\beta$ are linear transformation parameters, $\gamma$ is exponential transformation parameters. 
 We follow the setting of ~(\cite{lv2019attention}) so that they obey the following uniform distribution: $\alpha$ $\sim$ $\mathcal{U}(0.9,1)$,$\beta$ $\sim$ $\mathcal{U}(0.5,1)$,$\gamma$ $\sim$ $\mathcal{U}(1.5,5)$, respectively. By combining linear transformations with gamma functions, synthetic low-light images can be generated that resemble real low-light effects.
 \vspace{0.05cm}
 
 \noindent\textbf{Noise addition.} To expand the distribution of the data and improve the robustness of the network model to noise in low-light scenes, we further add noise to the low-light images. Some existing methods usually add Gaussian white noise directly to the original image, but this does not imitate the noise distribution in the real world well. 
 To introduce more realistic noise, we refer to the noise model on the RAW domain in ~(\cite{guo2019toward,brooks2019unprocessing}) map the image from the sRGB domain to the RAW domain through the ISP pipeline, and add shot and read noise to low-resolution, low-light images. 
 Shot noise is generally considered to follow a Poisson distribution, while read noise is generally considered to follow a Gaussian distribution with zero mean and fixed variance. 
 The combined noise distribution of the two can be approximated as a Gaussian distribution with zero mean.
 In this way, the additional shot and read noise can be expressed as $n(x)\sim N(0,\sigma^{2}(x)=x\cdot \sigma^{2}_s+\sigma^{2}_r)$. 
 $\sigma^{2}_s$ and $\sigma^{2}_r$ are the variances of shot noise and read noise, and obey the distribution: $\log{\sigma^{2}_s} \sim \mathcal{U}(\log{0.0001},\log{0.012})$ and $\log{\sigma^{2}_r}|\log{\sigma^{2}_s} \sim \mathcal{N}(2.18\cdot \log{\sigma^{2}_s}, 0.26)$, respectively.
 The image with shot and read noise added in this way can approximately obey the following heteroscedastic Gaussian distribution: $I_{noise} \sim \mathcal{N}(I_{raw},I_{in}\cdot \sigma^{2}_s + \sigma^{2}_r)$, where $I_{noise}$ is the additional noise image, $I_{raw}$ is the noise-free image. The model for adding noise through the ISP process is defined as
 $I_{raw}=F^{-1}_{wb}(F^{-1}_{DM}(F^{-1}_{cc}(F^{-1}_{crf}(I_{in}))))$ and
 $I_{noise}=F_{crf}(F_{cc}(F_{DM}(F_{wb}(I_{raw}+ n(x))))),$
 where $F_{crf}$,$F_{cc}$,$F_{DM}$,$F_{wb}$ and $F^{-1}_{wb}$,$F^{-1}_{DM}$,$F^{-1}_{cc}$,$F^{-1}_{crf}$ are the tone mapping function, color correction function, graphics function and White balance function and its corresponding inverse operation.
 Finally, the image is mapped from the RAW domain back to the sRGB domain through the inverse process of the above ISP operation to obtain the final low-light, low-resolution image with noise.

 \section{EXPERIMENTAL RESULTS} 
 In this section, firstly, the experimental settings are introduced. Furthermore, we make a series of comparison among our method and other state-of-art approaches on the proposed synthetic dataset and real-world dataset. Lastly, the ablation studies are conducted to prove the effectiveness of our proposed method.  
 \subsection{Experimental Settings}
 We use the real-word dataset RELLISUR~(\cite{aakerberg2021rellisur}) and the proposed synthetic dataset DFSR-LLE for training and testing. The RELLISUR dataset is currently the only real world dataset for low-light images super-resolution acquired by sensors. 
 We employ various data augmentation operations, i.e., random cropping, random rotation, random flip. 
 Based on the low-light low-resolution image, we perform histogram equalization to obtain the adjusted image, and then calculate the resolution- and illumination-invariant map, finally, we calculate the maximum gradient for the resolution-cum-invariant map and concat the above four images as input to the model. We use Adam as the optimizer of the model and set the hyperparameter as $(\beta_1,\beta_2)=(0.9,0.99)$ with the initial learning rate to $0.0001$ and batch size to $16$, and  warmup is performed to make our network training more stable. We conduct our experiments on an NVIDIA GeForce GTX 2080Titan GPU with the PyTorch 1.8.0 framework. In order to quantitatively compare the performance of the models, we choose three measurement algorithms: \textit{Peak Signal to Noise Ratio (PSNR)}~(\cite{chan1983hardware}) and \textit{Structural Similarity Index (SSIM)}~(\cite{wang2004image}). 
 Under the premise of fairness, all methods are retrained until the best performance is achieved.

 \begin{table*}[htb]
 	\renewcommand{\arraystretch}{1.0}
 	\caption{Quantitative comparison among different methods on RELLISUR Dataset. Notice that all methods are retrained until the best performance is achieved. The red and blue values denote the best and second best performance, respectively.} 
 	\setlength{\tabcolsep}{0.2mm}{
 		\begin{tabular}{|c|c|c|c|c|c|c|c|c|c|c|c|}
 			\hline 
 			\multirow{1}{*}{Scale} &Metrics &D-DBPN &EDSR &PAN &SRFBN &SwinIR &RUAS$\rightarrow$SwinIR &MIRNet &SRFormer &HAT &Ours \\
 			
 			\hline
 			\hline
 			\multirow{2}{*}{$\times$2}&PSNR &18.70 &18.38 &18.78 &18.42 &18.38 &17.15 &\textbf{\textcolor{blue}{21.05}} &19.55 &20.21 &\textbf{\textcolor{red}{23.40}}\\	 
 			&SSIM &0.682 &0.679 &0.693 &0.662 &0.640 &0.617 &\textbf{\textcolor{blue}{0.720}} &0.704 &0.719 &\textbf{\textcolor{red}{0.753}} \\ 
 			\hline
 			\multirow{2}{*}{$\times$4}&PSNR &17.96 &17.69 &18.10 &17.67 &17.07 &19.78 &19.34 &18.72 &\textbf{\textcolor{blue}{19.75}} &\textbf{\textcolor{red}{21.58}} \\
 			&SSIM &0.674 &0.679 &0.700 &0.665 &0.663 &0.662 &0.704 &0.705 &\textbf{\textcolor{blue}{0.715}} &\textbf{\textcolor{red}{0.742}} \\
 			\hline		
 		\end{tabular}
 	}
 	\label{tab:RELLISUR} 
 \end{table*}

\begin{figure*}[htbp]
	\centering 
	\begin{tabular}{c}
		\includegraphics[width=13cm]{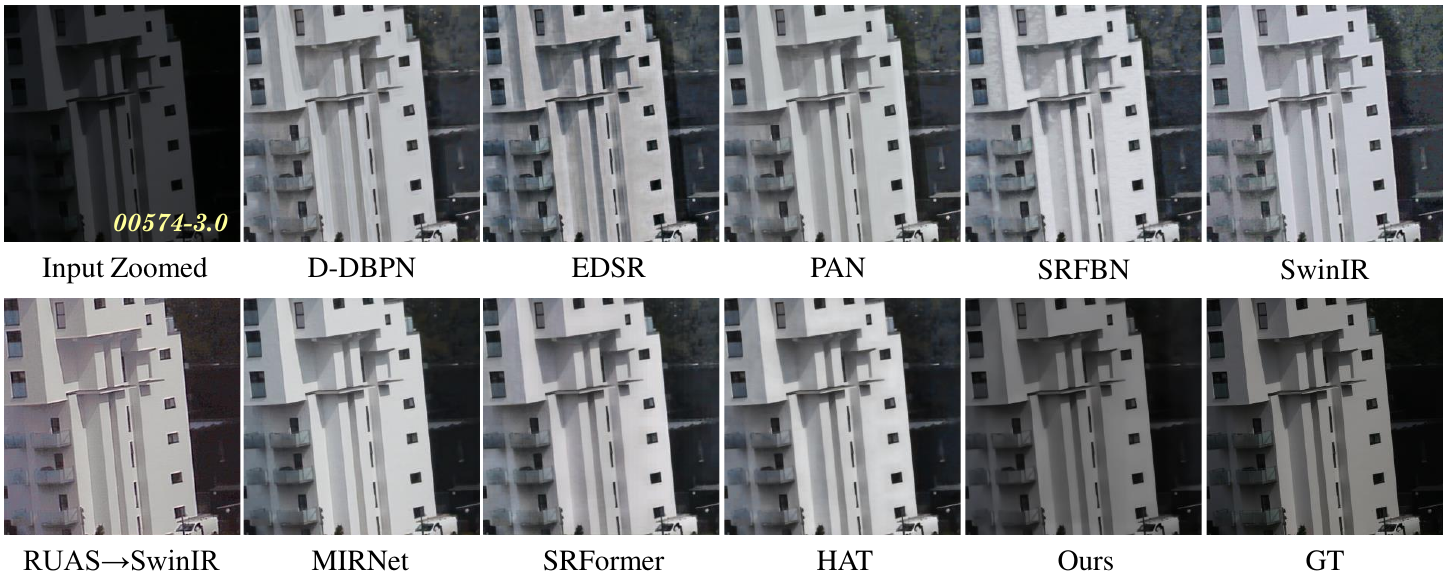}\\
	\end{tabular}%
	\caption{Simultaneous LLE and SR results of different algorithms and ours trained and tested on RELLISUR Dataset($\times2$). Our method does not over-brighten the image and does not add noise and artifacts. \textit{Zoom in for best view.} } 
	\label{fig:real_x2} 
\end{figure*}%

\begin{figure*}[htb]
	\centering 
	\begin{tabular}{c}
		\includegraphics[width=13cm]{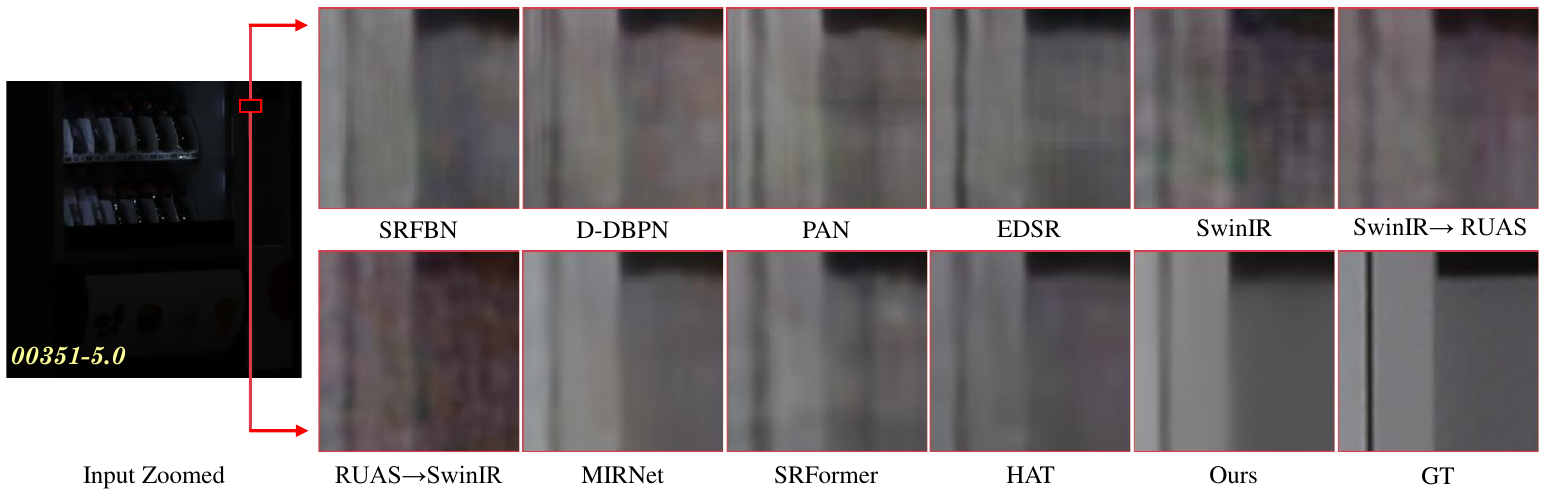}\\
	\end{tabular}%
	\caption{Simultaneous LLE and SR results of different algorithms and ours trained and tested on RELLISUR Dataset($\times2$). Our method maintains the texture details of an image in extremely dark environments without zooming in on the noise. } 
	\label{fig:real_x2_2} 
\end{figure*}%

\begin{figure*}[htb]
	\centering 
	\begin{tabular}{c}
		\includegraphics[width=13cm]{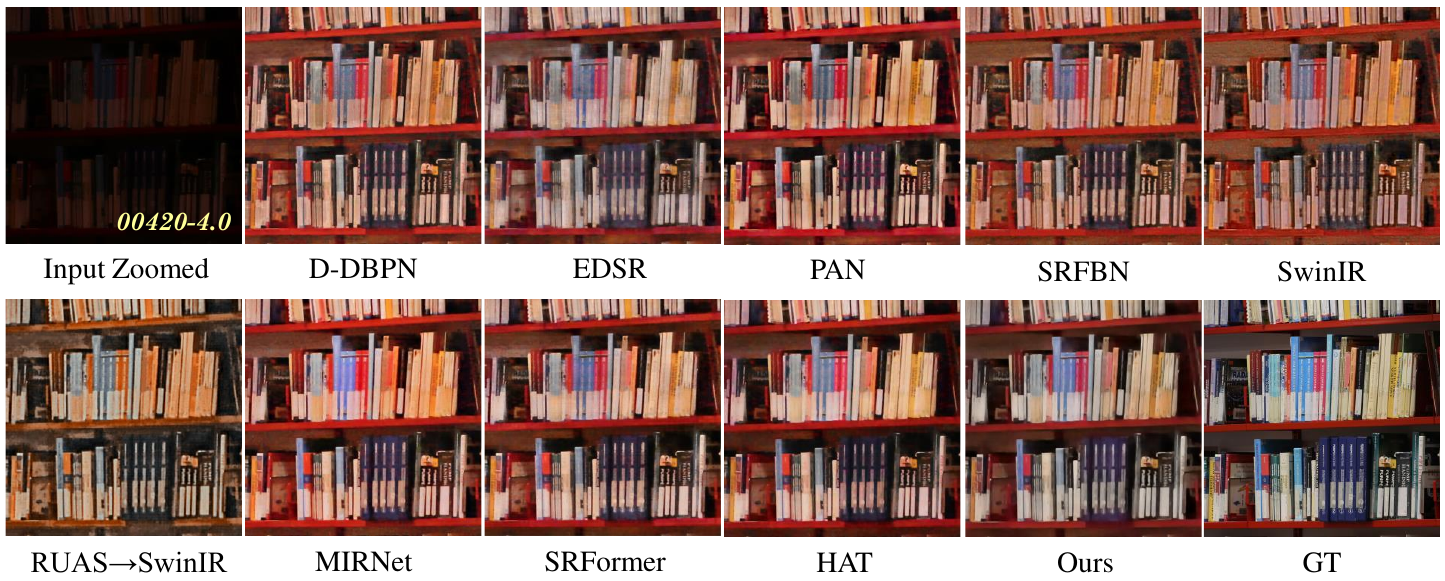}\\
	\end{tabular}%
	\caption{Simultaneous LLE and SR results of different algorithms and ours trained and tested on RELLISUR Dataset($\times4$). Our method does not produce severe artifacts and color deviations and is closer to ground truth.. \textit{Zoom in for best view.} } 
	\label{fig:real_x4} 
\end{figure*}%

\begin{figure*}[htb]
	\centering 
	\begin{tabular}{c}
		\includegraphics[width=13cm]{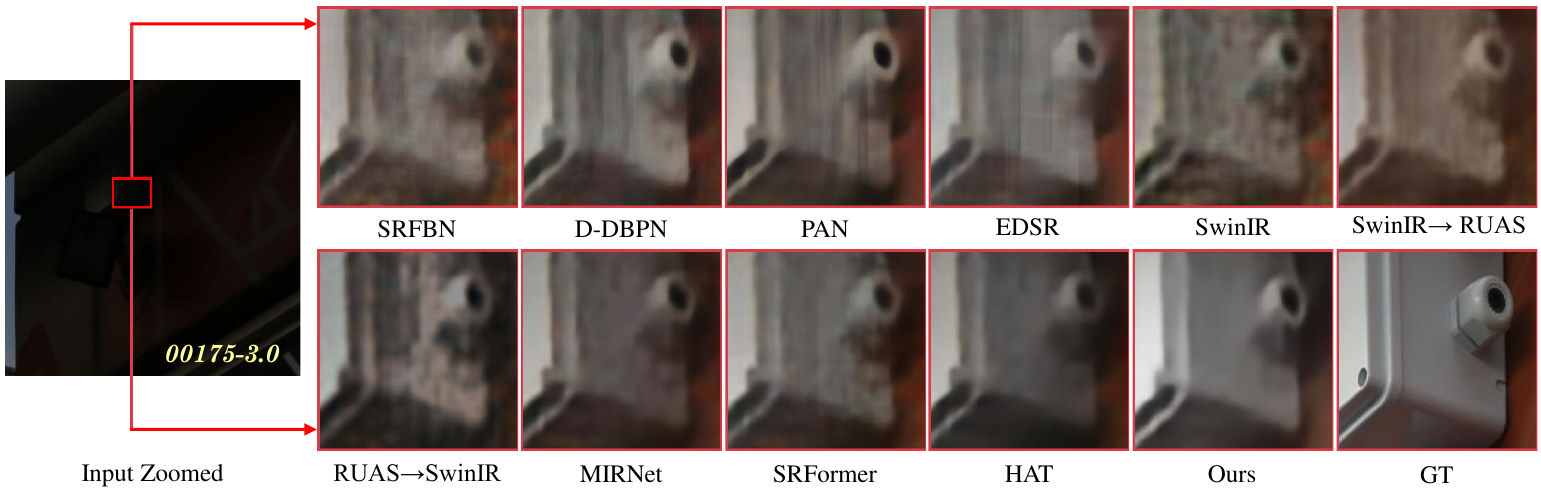}\\
	\end{tabular}%
	\caption{Simultaneous LLE and SR results of different algorithms and ours trained and tested on RELLISUR Dataset($\times4$). Compared with other methods, our method can effectively suppress the problems of noise amplification and artifacts. \textit{Zoom in for best view.} } 
	\label{fig:real_x4_2} 
\end{figure*}%

\begin{table*}[htb]
	\renewcommand{\arraystretch}{1.0}
	\caption{Quantitative comparison among different methods on DFSR-LLE Dataset.  Notice that all methods are retrained until the best performance is achieved. The red and blue values denote the best and second best performance, respectively.} 
	\setlength{\tabcolsep}{1mm}{
		\begin{tabular}{|c|c|c|c|c|c|c|c|c|c|}
			\hline 
			\multirow{1}{*}{Scale} &Metrics &MSRResNet &EDSR &PAN &SwinIR &MIRNet &srformer &HAT &Ours \\
			
			\hline
			\hline
			\multirow{2}{*}{$\times$2}&PSNR &19.76 &19.67 &20.04 &20.44 &20.70 &20.76 &\textbf{\textcolor{blue}{21.01}} &\textbf{\textcolor{red}{21.53}}\\	
			&SSIM &0.718 &0.717 &0.728 &0.745 &0.721 &\textbf{\textcolor{blue}{0.749}} &0.744 &\textbf{\textcolor{red}{0.753}} \\
			\hline
 			\multirow{2}{*}{$\times$4}&PSNR &19.02 &18.94 &19.24 &19.63 &19.33 &20.10 &\textbf{\textcolor{blue}{20.15}} &\textbf{\textcolor{red}{20.39}} \\
			&SSIM &0.626 &0.619 &0.630 &0.638 &0.601 &\textbf{\textcolor{blue}{0.640}} &0.638 &\textbf{\textcolor{red}{0.642}} \\
			\hline		
		\end{tabular}
	}
	\label{tab:DFSR} 
\end{table*}

\begin{figure*}[htb]
	\centering 
	\begin{tabular}{c}
		\includegraphics[width=13cm]{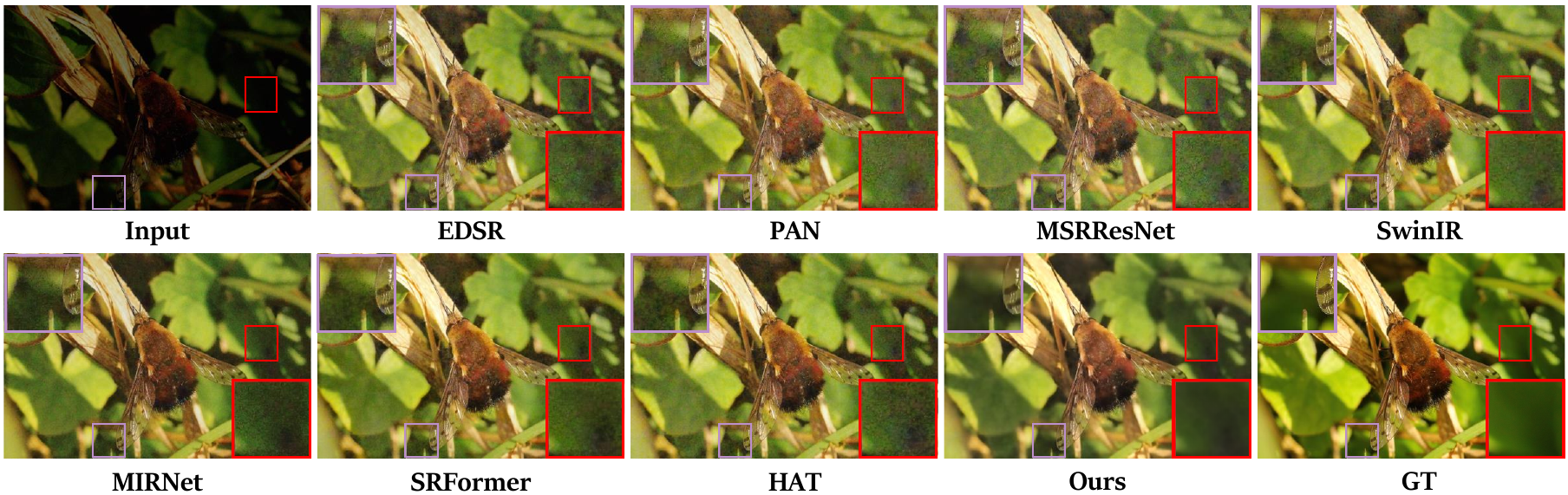}\\
	\end{tabular}%
	\caption{Visual results of state-of-the-art methods and ours on DFSR-LLE Dataset($\times2$). Our method can restore brightness and color details perfectly and are able to remove the noise. \textit{Zoom in for best view.}} 
	\label{fig:DFSR_x2} 
\end{figure*}%

\begin{figure*}[htb]
	\centering 
	\begin{tabular}{c}
		\includegraphics[width=13cm]{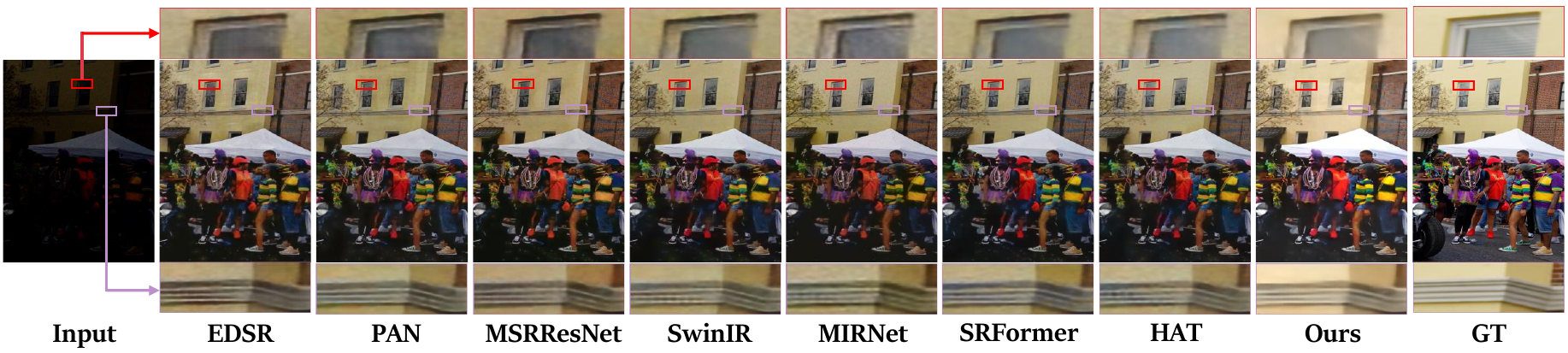}\\
	\end{tabular}%
	\caption{Visual results of state-of-the-art methods and ours on DFSR-LLE Dataset($\times4$). Our algorithm recovers a high-quailty image with clearer structures and details. \textit{Zoom in for best view.} } 
	\label{fig:DFSR_x4} 
\end{figure*}%
 
 \subsection{Comparisons with State-of-the-art Methods}
 
 \noindent\textbf{Evaluation on the real-world dataset.}
 We first compare the visual quality of LoLiSRFlow with several recent competing methods on the RELLISUR dataset ~(\cite{aakerberg2021rellisur}). These methods include separate SR methods(MSRFBN~(\cite{li2019feedback}), D-DBPN~(\cite{haris2018deep}), EDSR~(\cite{lim2017enhanced}), PAN~(\cite{zhao2020efficient}), SwinIR~(\cite{liang2021swinir}), SRFormer~(\cite{zhou2023srformer}) and HAT~(\cite{chen2023activating})), concatenation of LLE methods with upsampling modules (MIRNet~\cite{han2020mirnet}), and sequential concatenation of SR and LLE methods (RUAS$\rightarrow$SwinIR). Results are demonstrated in Table~\ref{tab:RELLISUR}, Figure~\ref{fig:real_x2}, Figure~\ref{fig:real_x2_2}, Figure~\ref{fig:real_x4} and Figure~\ref{fig:real_x4_2}. 
 The visualization shows $2\times$, $4\times$ full image results and patch image results.
 It is obvious that both the sequential cascade approach, the direct use of a super-resolution depth model and the combination of LLE and up-sampling suffer from overall over-brightening, color deviation, noise amplification and poor local detail recovery.
 Compared to the above methods, the images recovered by the proposed LoLiSRFlow are closer to the ground truth in terms of color and detail and do not amplify noise, as evidenced by the highest PSNR and SSIM metrics in double and quadruple experiments.
 
 \noindent\textbf{Evaluation on the proposed synthetic dataset.}
 We secondly compare the visual quality of LoLiSRFlow with several state-of-the-art SR methods on the proposed synthetic dataset. The methods we compare are EDSR~(\cite{lim2017enhanced}), MSRFBN~(\cite{li2019feedback}), PAN~(\cite{zhao2020efficient}), MIRNet~(\cite{han2020mirnet}), SwinIR~(\cite{liang2021swinir}), SRFormer~(\cite{zhou2023srformer}) and HAT~(\cite{chen2023activating}). All state-of-the-art SR methods and our method also have been trained and tested on the proposed synthetic dataset. As shown in Table~\ref{tab:DFSR}, our method achieves the best quantitative results. Visual results are demonstrated in Figure~\ref{fig:DFSR_x2} and Figure~\ref{fig:DFSR_x4}. 
 As shown by the purple and red patches in Figure~\ref{fig:DFSR_x2}, the rest of the state-of-the-art methods, despite brightening and restoring details, also amplify the noise, compared to our proposed method which effectively suppresses the noise while incrementing and improving the resolution.
 In Figure~\ref{fig:DFSR_x4}, it can be observed that the other methods all suffer from insufficient brightening, as well as introducing artifacts and amplifying noise to varying degrees.
 In contrast, our method achieves better results.
 Our method significantly outperforms all the other competitors achieving more good perceptual quality by better suppressing the artifacts and revealing image details. The results generated by our method are less noise and better color consistency, which illustrates that LoLiSRFlow can restore brightness and points perfectly.
 
 \begin{table*}[htb]
 	\centering
 	\caption{Quantitative comparison for different architectures of encoder and losses on the real dataset. $\uparrow(\downarrow)$ denotes that, larger (smaller) values lead to better quality. Notice that all methods are retrained until the best performance is achieved. \cmark and \xmark~ represent whether this module is used or not, respectively.} 
 	\setlength{\tabcolsep}{0.1 mm}{
 		\begin{tabular}{cccccccccc}
 			\toprule
 			\multicolumn{1}{c}{Training Dataset}& \multicolumn{2}{c}{Encoder architecture} & \multirow{2}{*}{Invertible flow} &\multicolumn{2}{c}{Loss}&\multicolumn{2}{c}{$\times2$}&\multicolumn{2}{c}{$\times4$} \vspace{0.001in}\\		
 			\footnotesize  $(-2.5\rightarrow-5.0)$ EV & RRDB & MRPT &  & $L_1$ & $L_{nll}$ & PSNR $\uparrow$ & SSIM $\uparrow$& PSNR $\uparrow$ & SSIM $\uparrow$ \\
 			\midrule 
 			$(-2.5\rightarrow-5.0)$ EV &\cmark &\xmark &\cmark &\cmark &\cmark &21.02 &0.681 &19.95 &0.700\\ 
 			$(-2.5\rightarrow-5.0)$ EV &\xmark &\cmark &\cmark &\xmark &\cmark &22.94 &0.764 &20.75 &0.739\\
 			Only $-3.0$EV              &\xmark &\cmark &\cmark &\cmark &\cmark &23.03 &0.747 &20.77 &0.729\\ 
 			$(-2.5\rightarrow-5.0)$ EV &\xmark &\cmark &\cmark &\cmark &\cmark &23.41 &0.783 &21.57 &0.768\\
 			\bottomrule
 		\end{tabular}
 	}
 	\label{tab:ablation}  
 \end{table*}
 
 \begin{figure}[htb]
 	\begin{center}
 		\begin{tabular}{c@{\extracolsep{0.1em}}c@{\extracolsep{0.1em}}c@{\extracolsep{0.1em}}c@{\extracolsep{0.1em}}c@{\extracolsep{0.1em}}c@{\extracolsep{0.1em}}}
 			\hspace{-0.1cm}\includegraphics[height=1.8cm,width=1.63cm,trim=0 10 40 0,clip]{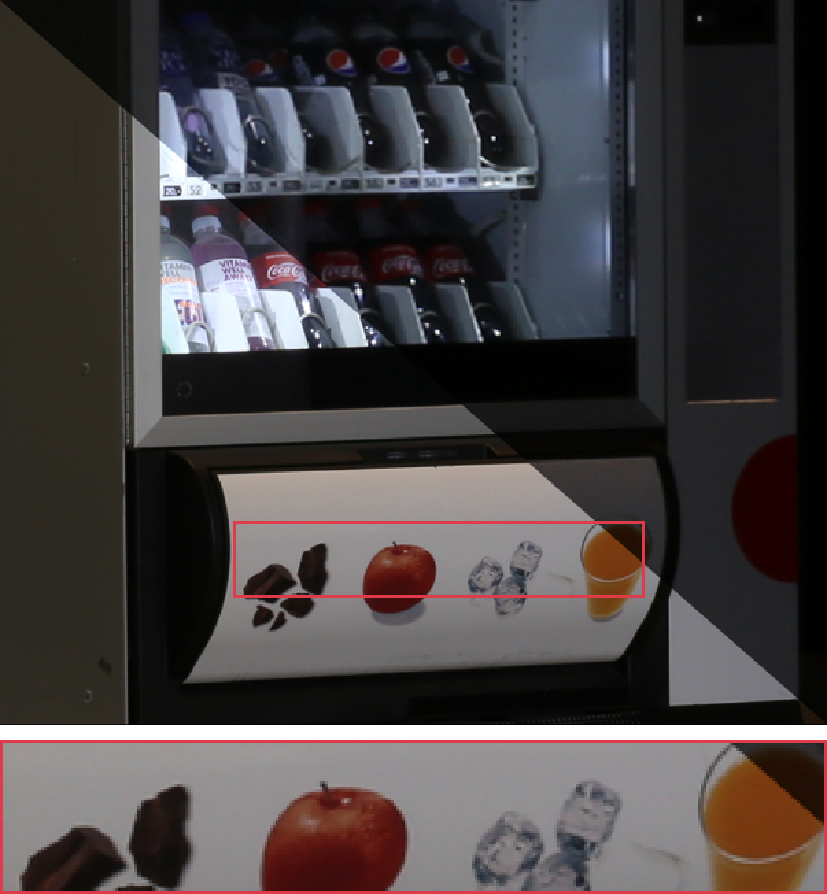}&			
 			\hspace{-0.1cm}\includegraphics[height=1.8cm,width=1.63cm,trim=0 10 40 0,clip]{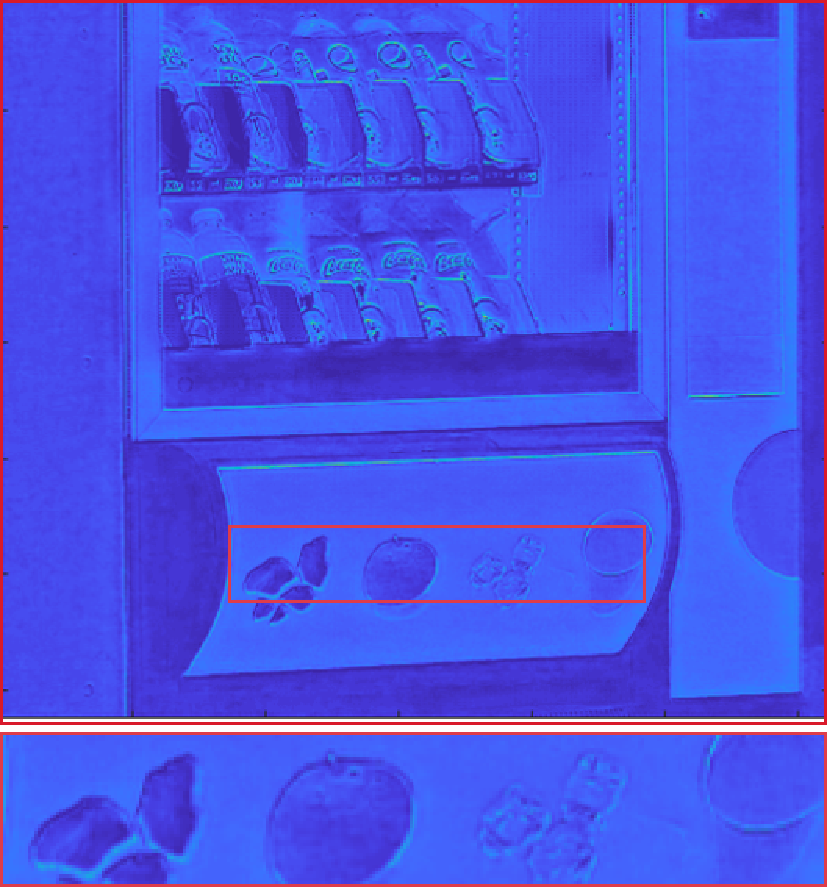}
 			&\hspace{-0.1cm}\includegraphics[height=1.8cm,width=1.63cm,trim=0 10 40 0,clip]{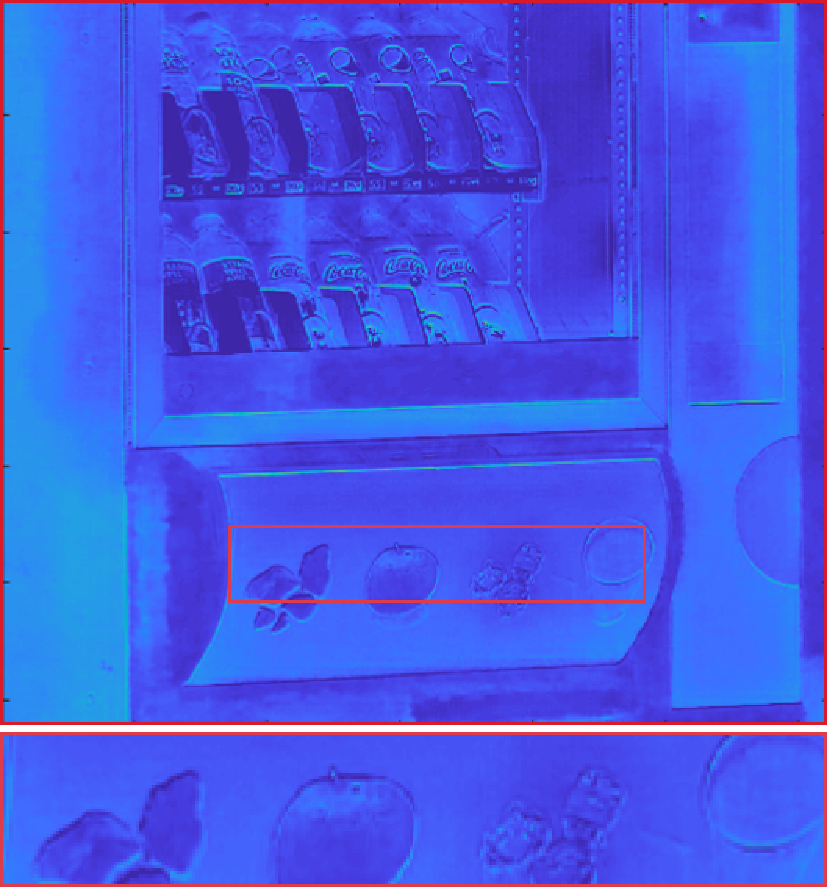}
 			&\hspace{-0.1cm}\includegraphics[height=1.8cm,width=1.63cm,trim=0 10 40 0,clip]{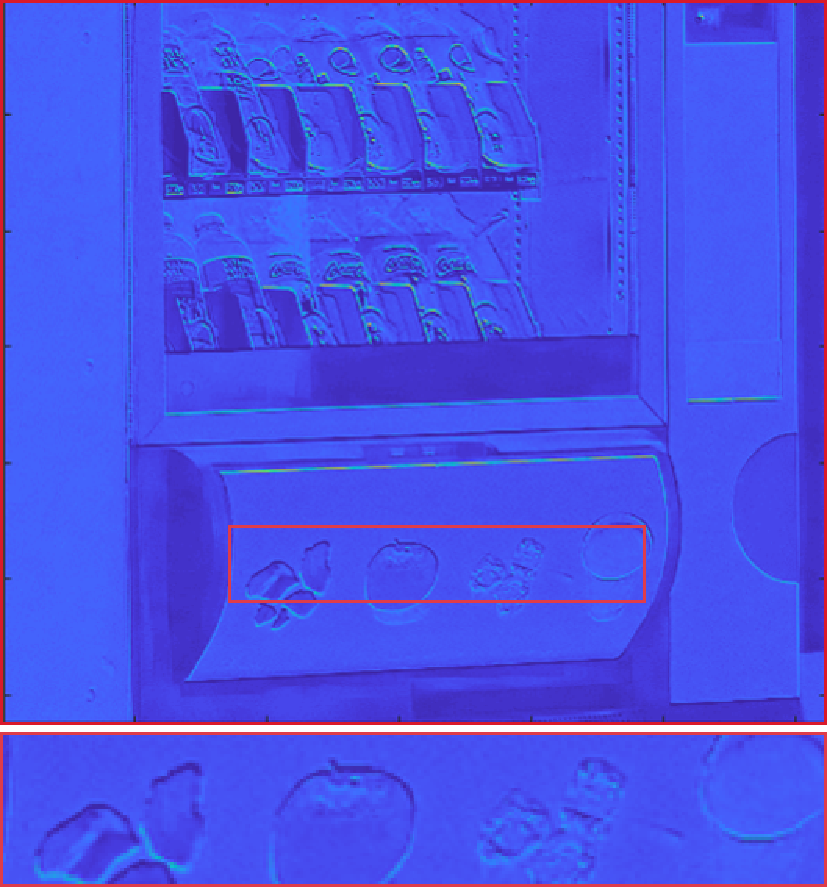}&
 			\hspace{-0.1cm}\includegraphics[height=1.8cm,width=1.63cm,trim=0 10 40 0,clip]{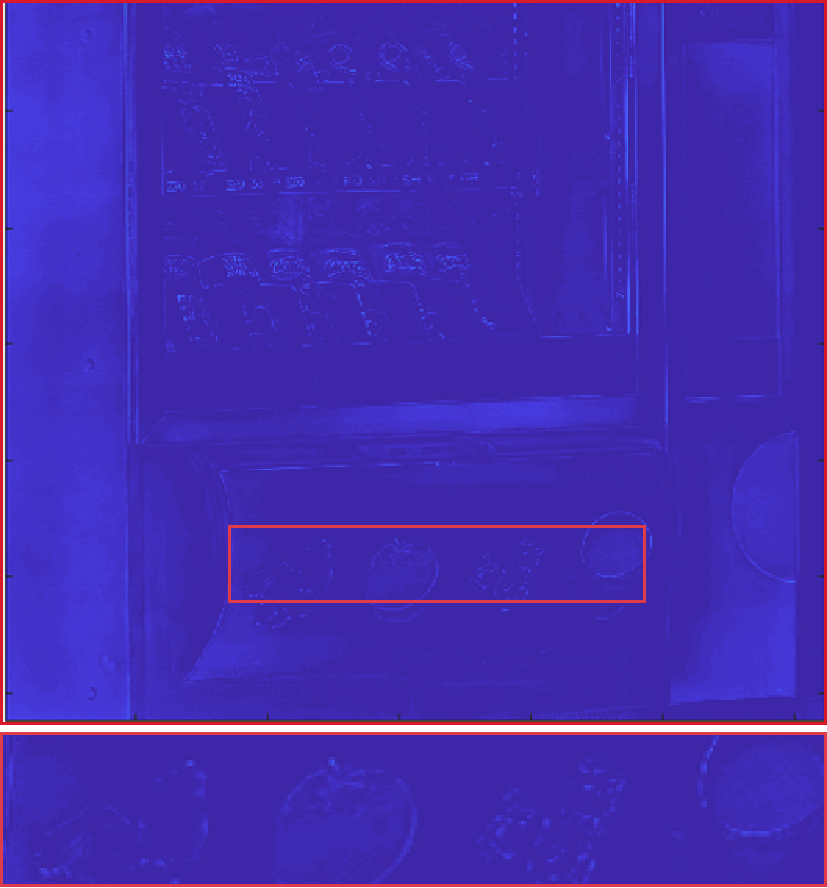}&
 			\hspace{-0.1cm}\includegraphics[height=1.8cm,width=0.3cm]{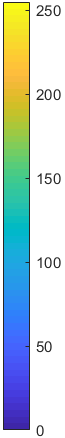}\\
 			\footnotesize Input  & \footnotesize ESRGAN  & \footnotesize SwinIR  &\footnotesize RealESRGAN &\footnotesize Ours &\\
 		\end{tabular}
 	\end{center}
 	\vspace{-0.4cm}
 	\caption{Comparison results of pixel error maps with GAN methods and transformer-based methods.  It indicates our generated images are friendlier in terms of detail recovery.} 
 	\label{fig:error_map}
 \end{figure}
 
 \noindent\textbf{Comparisons on pixel error heatmaps.}
 We also visualize the pixel error between the predictions and the ground truth in the form of a heatmap. As shown in Figure~\ref{fig:error_map}, our proposed method has more advantages compared with the three methods of ESRGAN~(\cite{wang2018esrgan}), RealESRGAN~(\cite{wang2021real}) and SwinIR~(\cite{liang2021swinir}). This demonstrates that the introduction of a multi-scale parallel transformer structure and a reversible normalized flow model can capture global features and high-frequency details more comprehensively, and restore the details in the picture well.  
 
 \begin{figure}[htb]
 	\begin{center}
 		\begin{tabular}{c@{\extracolsep{0.1em}}c@{\extracolsep{0.1em}}c@{\extracolsep{0.1em}}c@{\extracolsep{0.1em}}}			
 			\includegraphics[height=1.8cm,width=2.01cm,trim=0 0 40 40,clip]{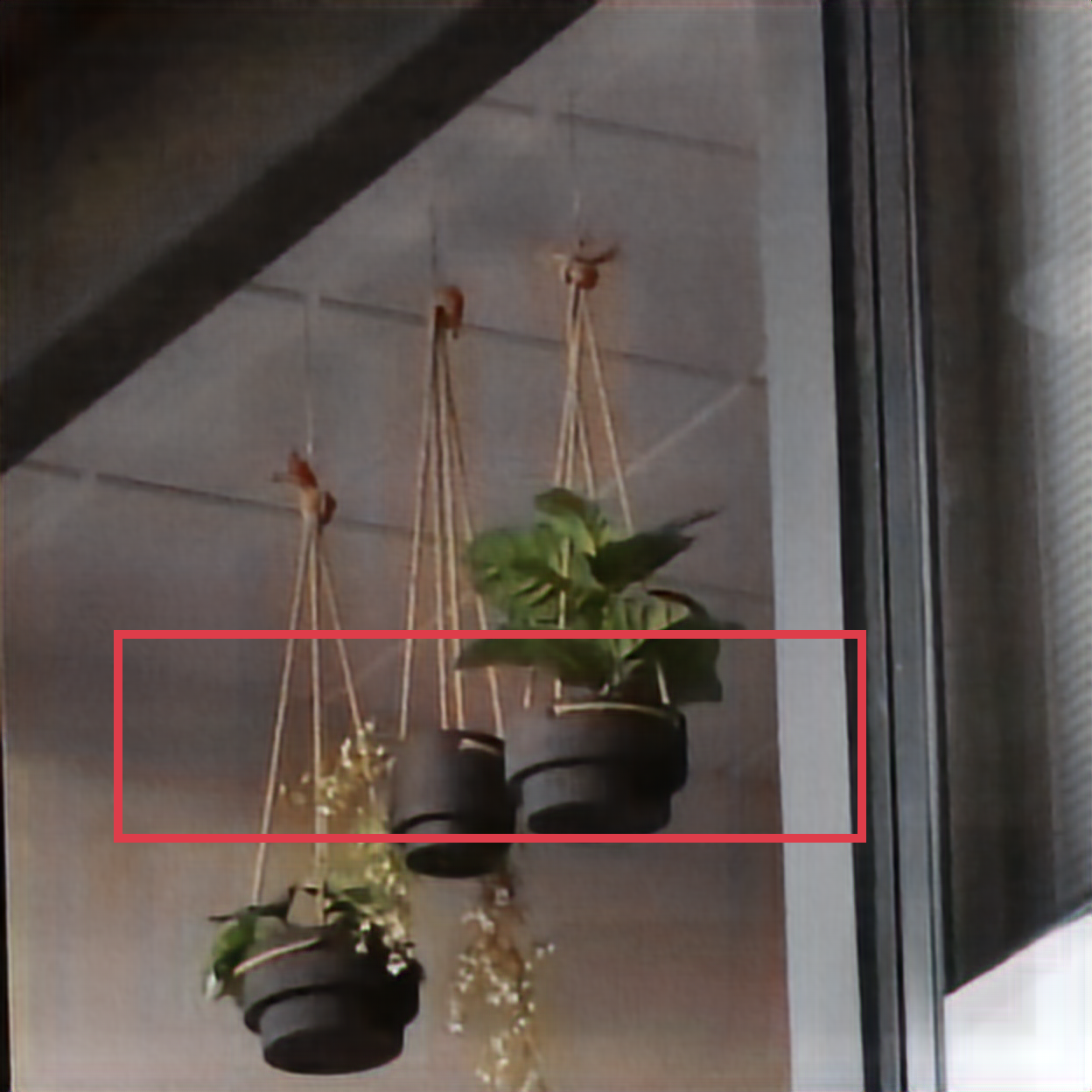}
 			&\includegraphics[height=1.8cm,width=2.01cm,trim=0 0 40 40,clip]{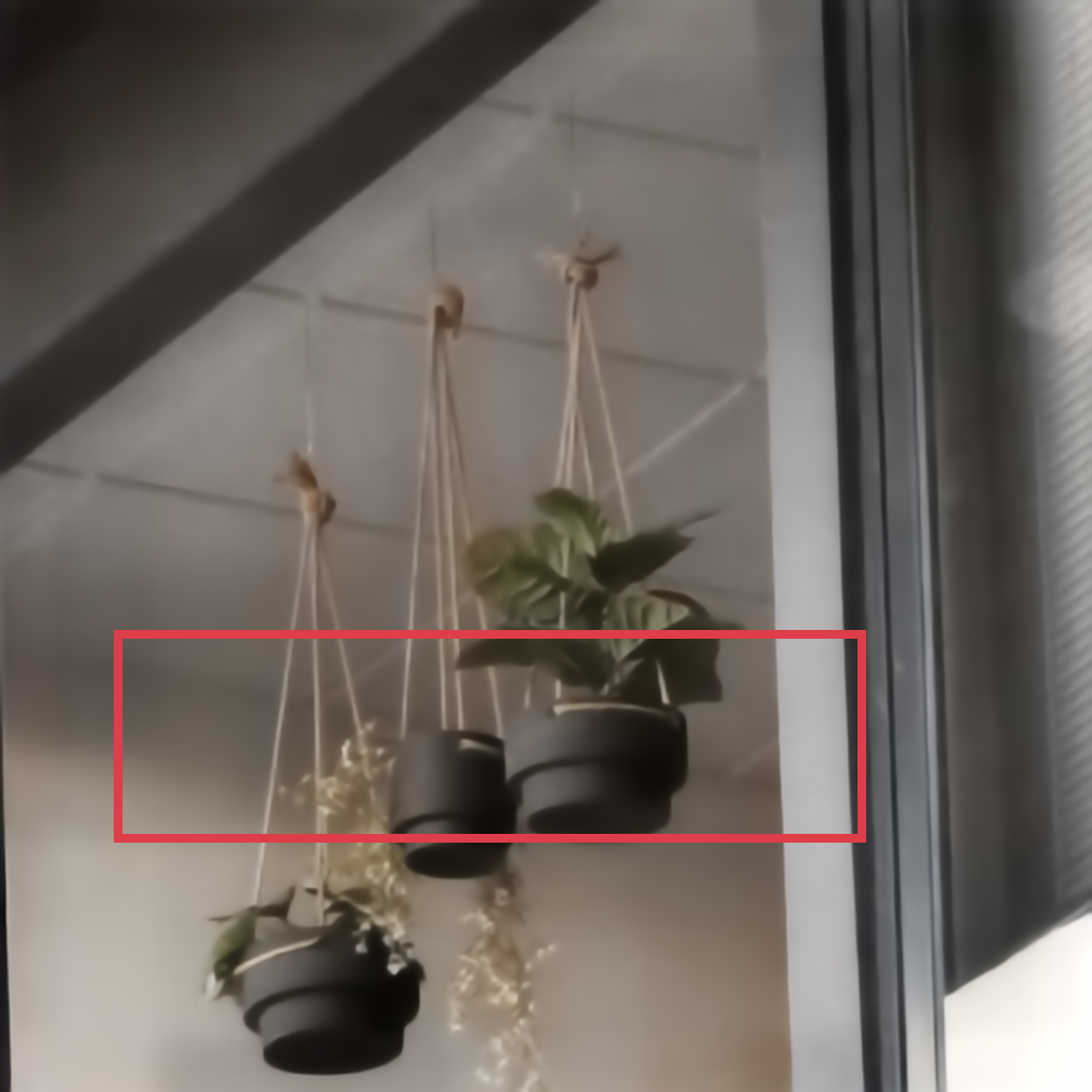}
 			&\includegraphics[height=1.8cm,width=2.01cm,trim=0 0 40 40,clip]{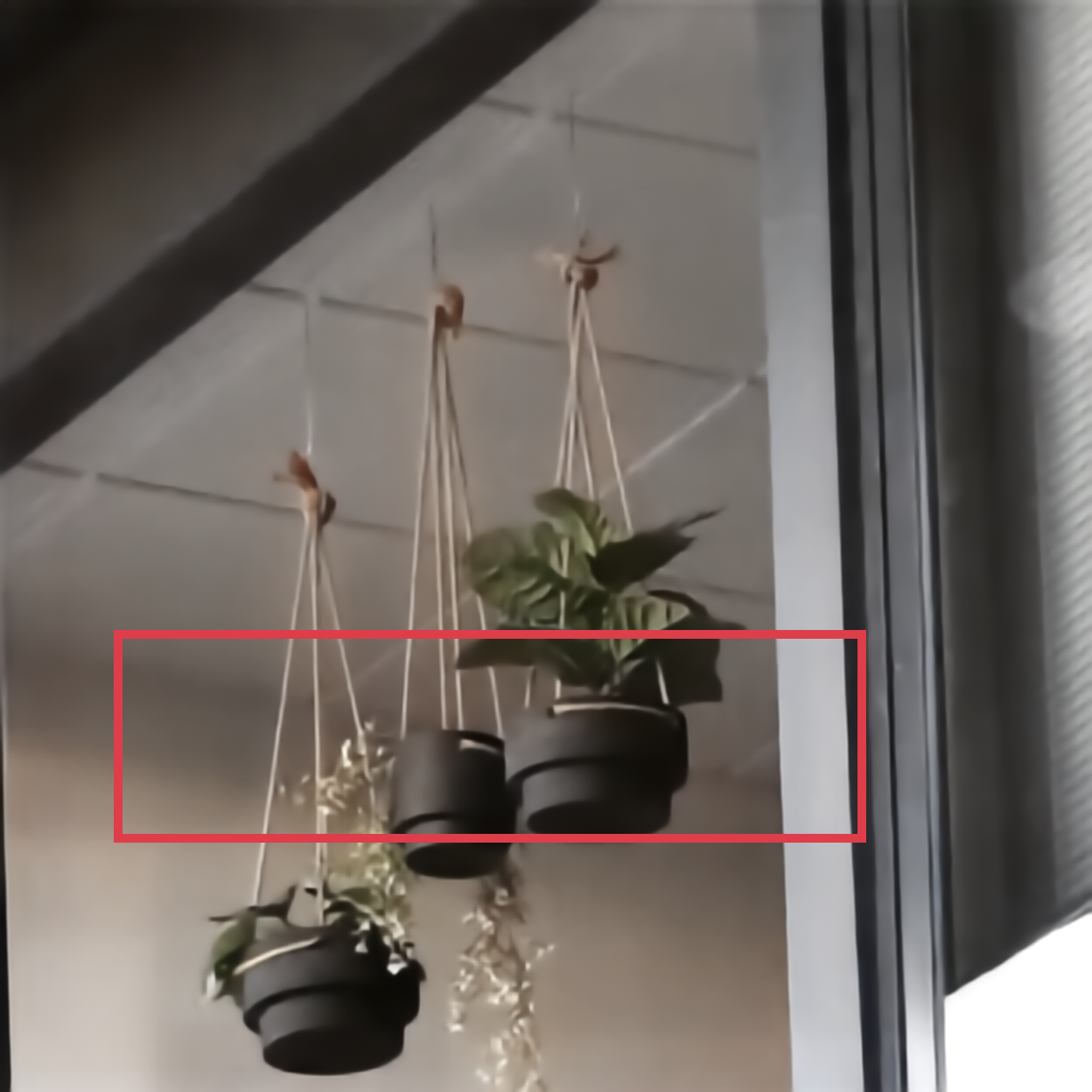}
 			&\includegraphics[height=1.8cm,width=2.01cm,trim=0 0 40 40,clip]{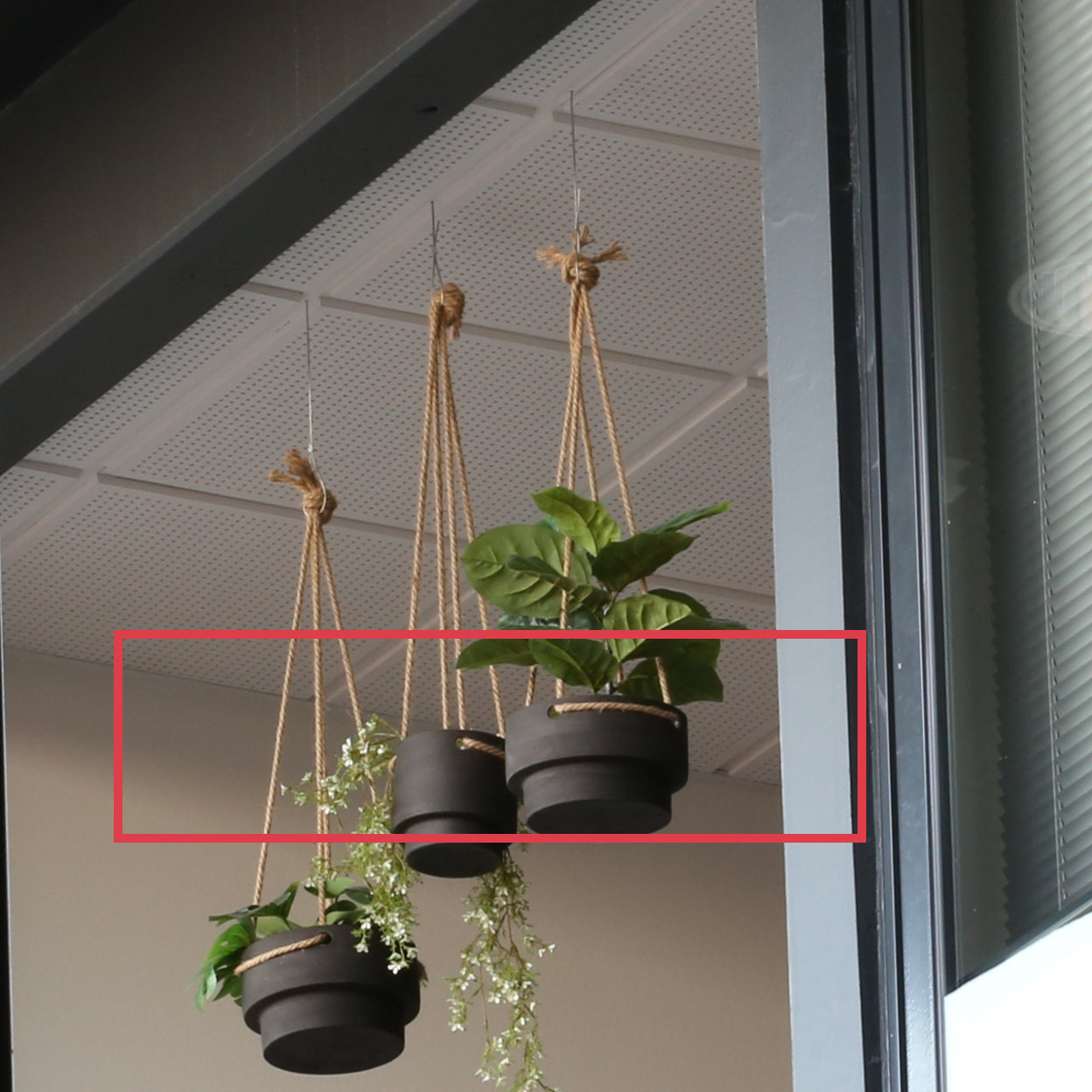}\\
 			\includegraphics[height=0.6cm,width=2.01cm,trim=0 0 40 40,clip]{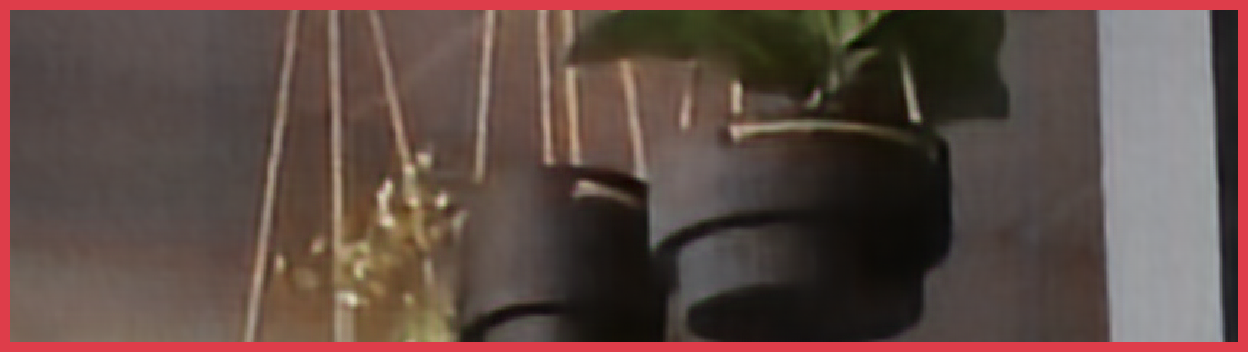}
 			&\includegraphics[height=0.6cm,width=2.01cm,trim=0 0 40 40,clip]{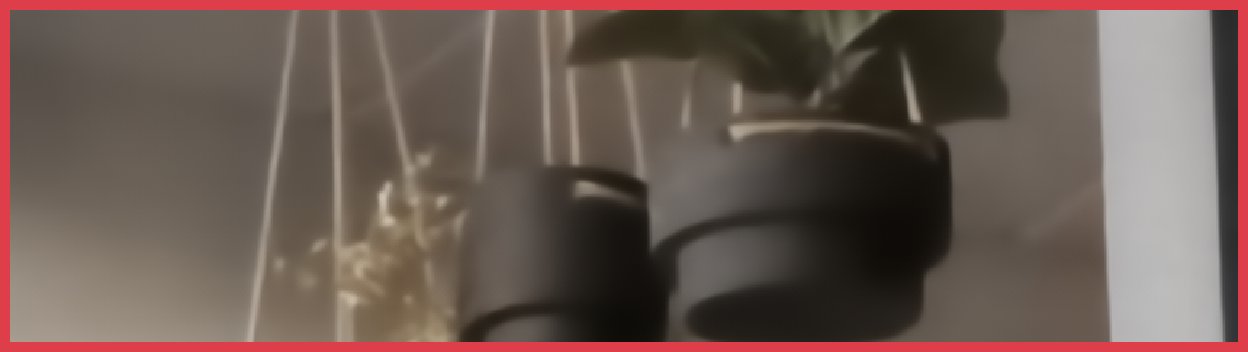}
 			&\includegraphics[height=0.6cm,width=2.01cm,trim=0 0 40 40,clip]{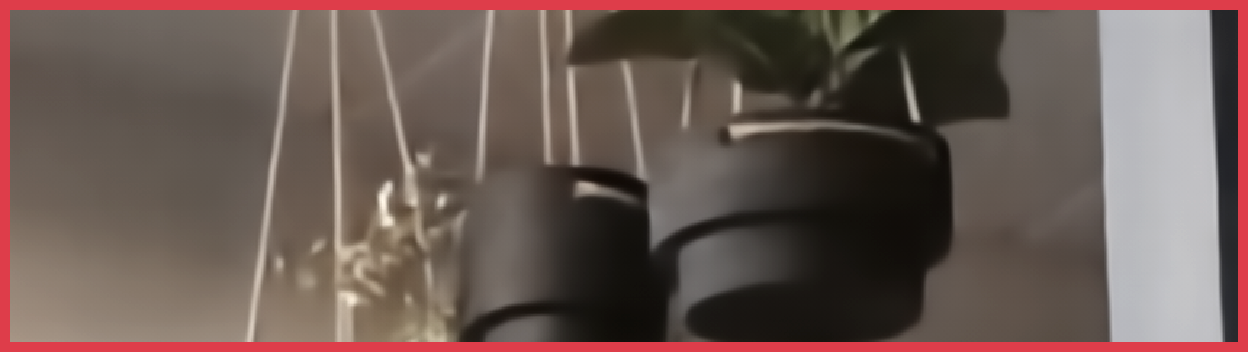}
 			&\includegraphics[height=0.6cm,width=2.01cm,trim=0 0 40 40,clip]{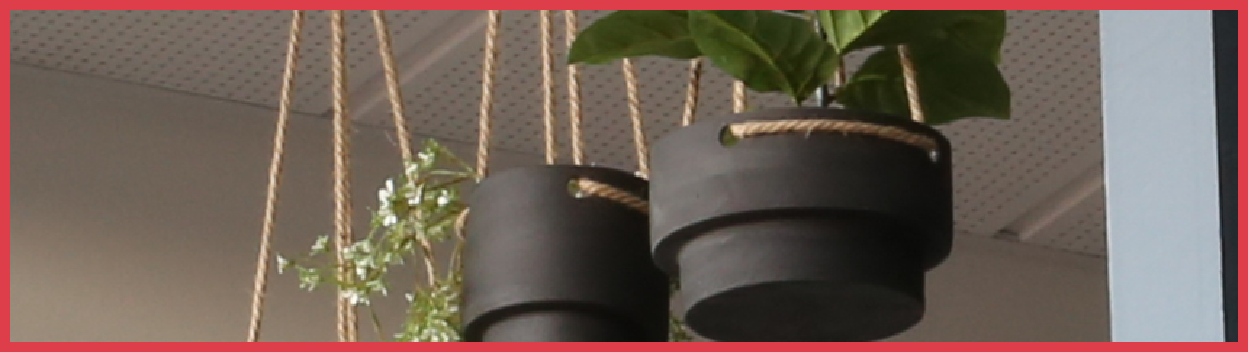}\\
 			\footnotesize $(20.38,0.76)$  & \footnotesize $(24.61,0.79)$  &\footnotesize $(25.38,0.80)$ &\footnotesize (PSNR, SSIM)\\
 			\footnotesize $\mathcal{L}_{1}$  & \footnotesize $\mathcal{L}_{nll}$  &\footnotesize $\mathcal{L}_{total}$ &\footnotesize Ground Truth\\
 		\end{tabular}
 	\end{center}  
 	\caption{Comparison using different Loss functions. Combining the two loss functions can guarantee to recover of more image details and keep content and color consistency. }
 	\label{fig:loss_compare}
 \end{figure}
 
 \subsection{Ablation study}
 \noindent\textbf{Effectivenesses of the model architecture:}
 To verify the effectiveness of our proposed transformer-based cross-scale conditional encoder, we train and test on the RELLISUR dataset with RRDB and the proposed encoder as baselines. As shown in Table~\ref{tab:ablation}, the transformer-based cross-scale conditional encoder has better performance under the supervision of same loss training.

\begin{figure}[htb]
	\begin{center}
		\begin{tabular}{c@{\extracolsep{0.2em}}c@{\extracolsep{0.2em}}c@{\extracolsep{0.2em}}}			
			\includegraphics[height=1.8cm,width=2.6cm,trim=0 0 40 40,clip]{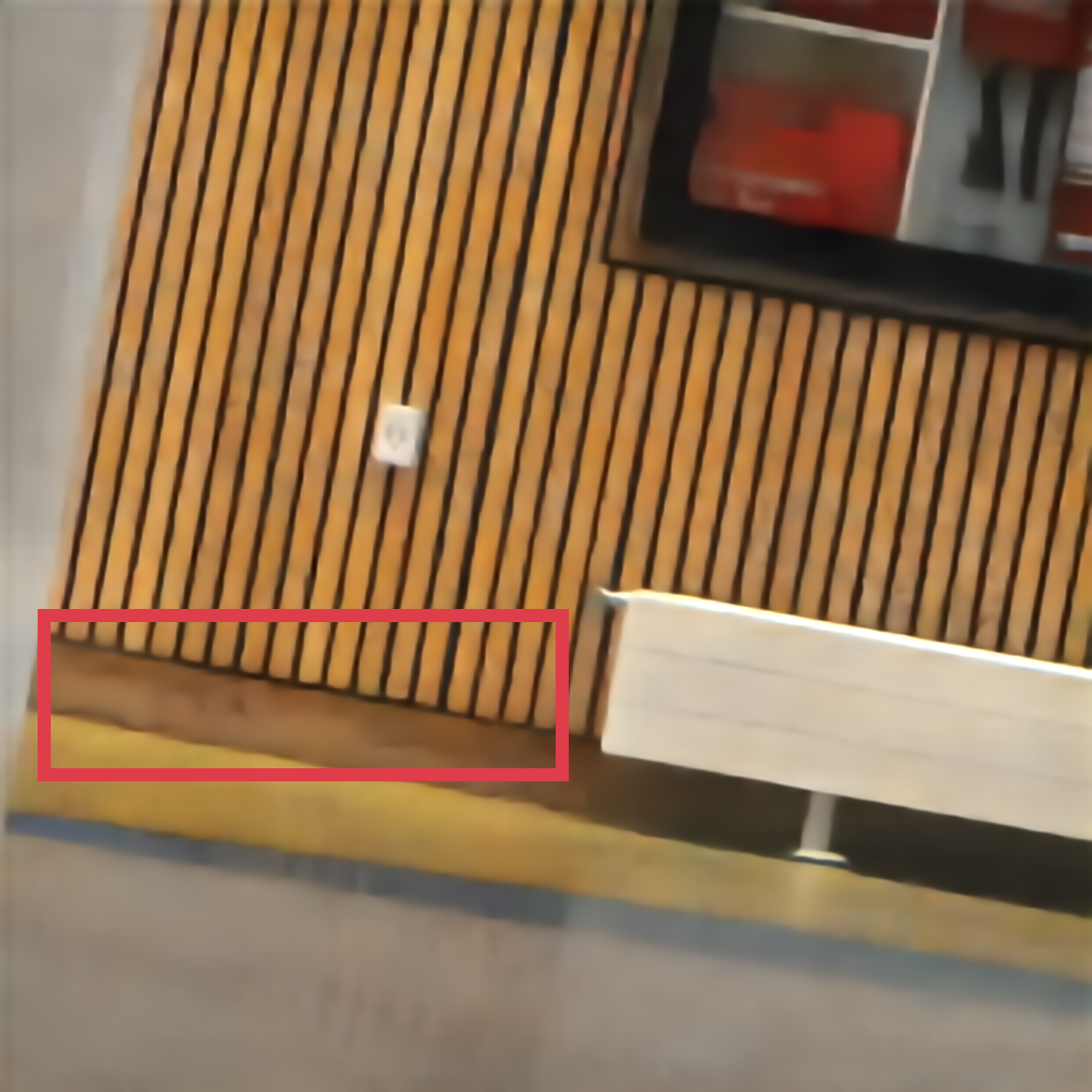}
			&\includegraphics[height=1.8cm,width=2.6cm,trim=0 0 40 40,clip]{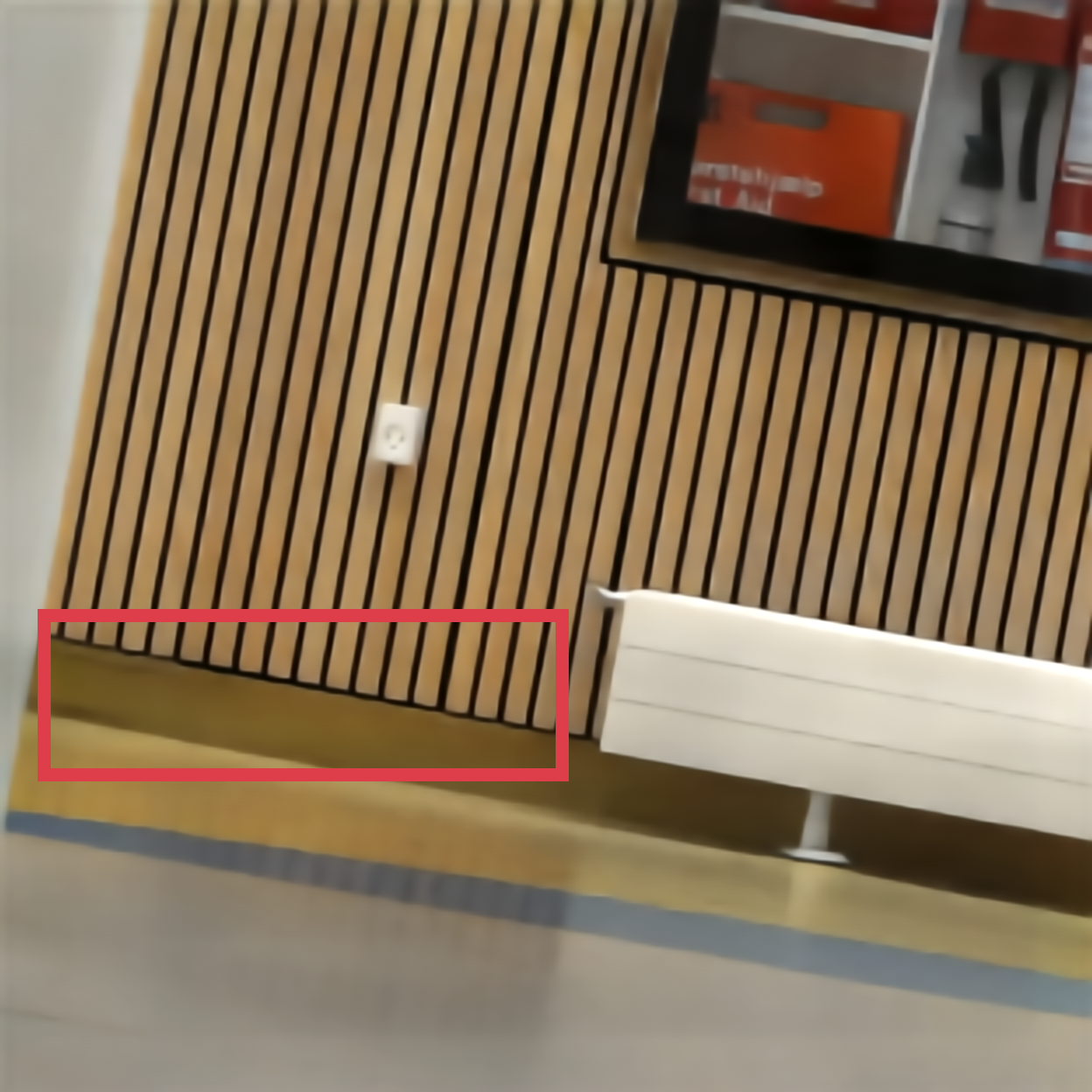}
			&\includegraphics[height=1.8cm,width=2.6cm,trim=0 0 40 40,clip]{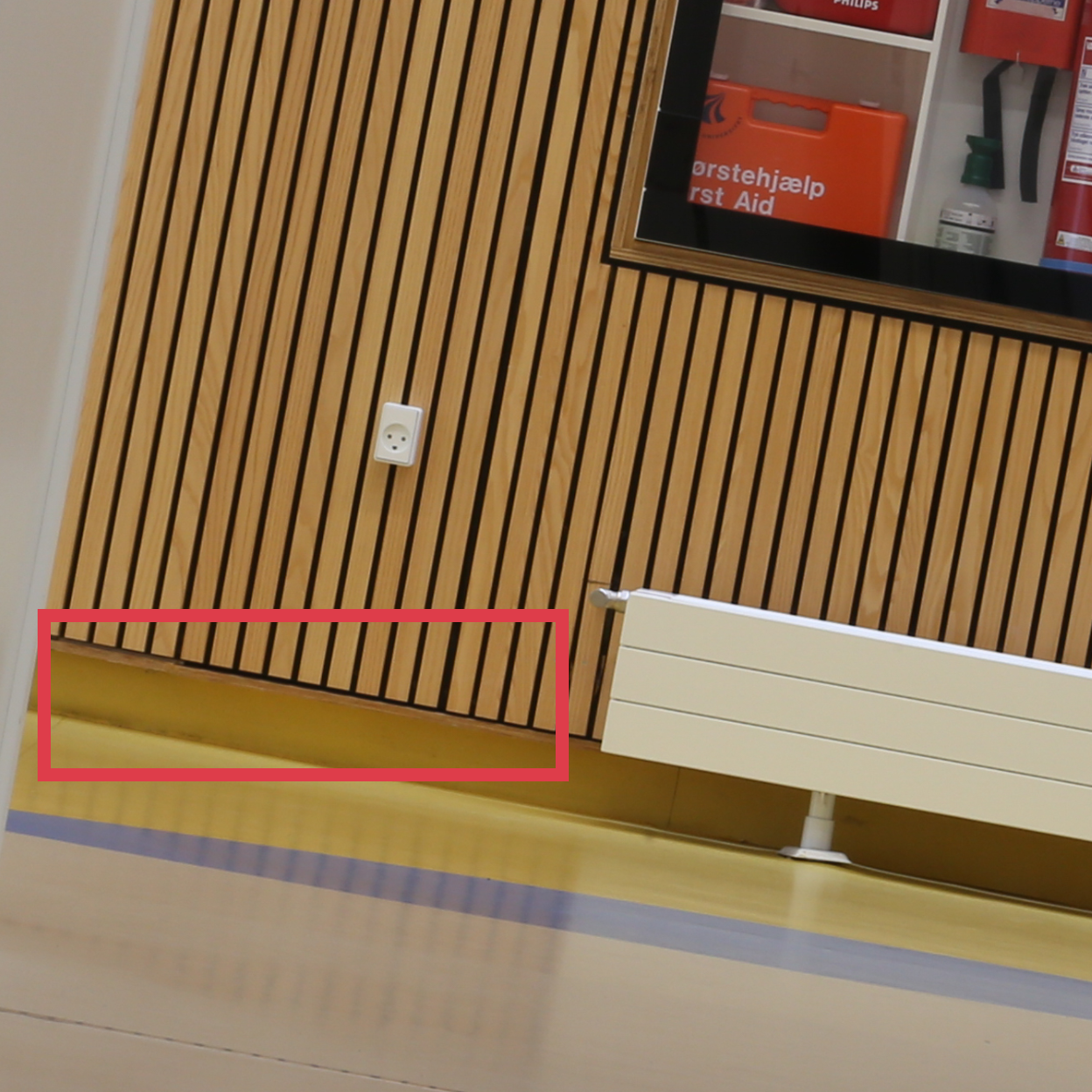}\\
			\includegraphics[height=0.6cm,width=2.6cm,trim=0 0 0 0,clip]{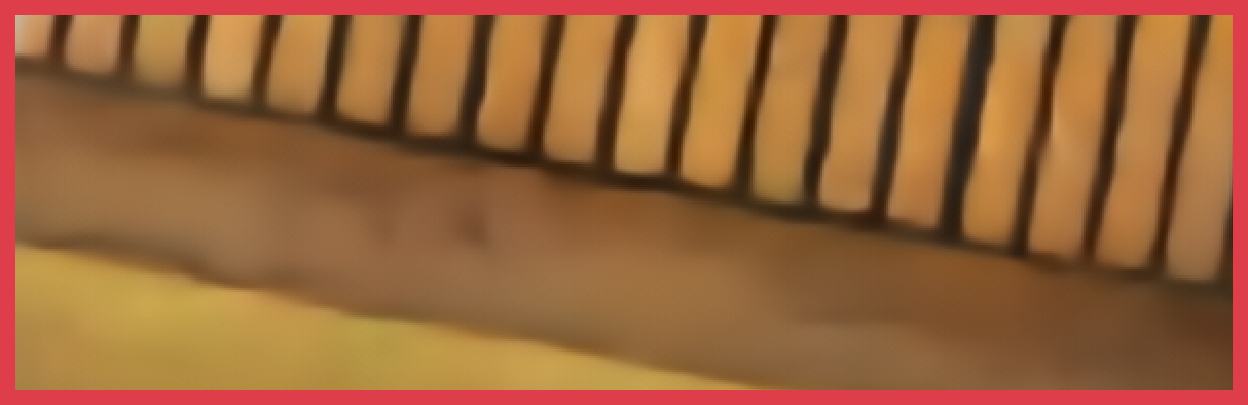}
			&\includegraphics[height=0.6cm,width=2.6cm,trim=0 0 0 0,clip]{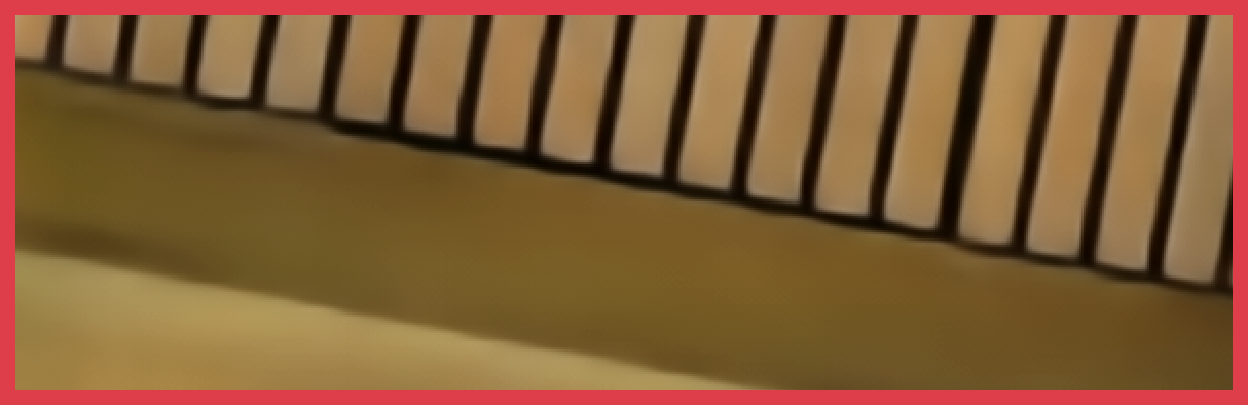}
			&\includegraphics[height=0.6cm,width=2.6cm,trim=0 0 0 0,clip]{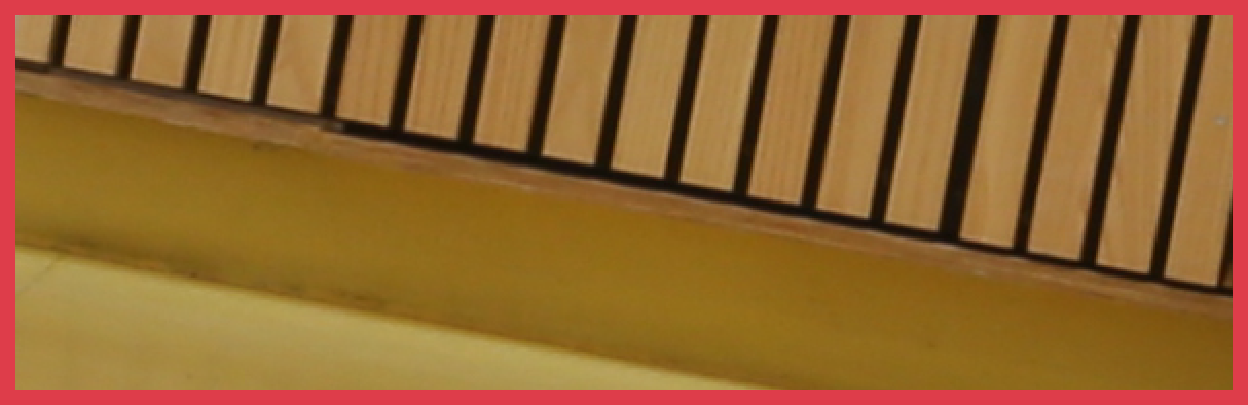}\\
			\footnotesize $(21.68,0.73)$  & \footnotesize $(22.13, 0.82)$   &\footnotesize (PSNR, SSIM) \\
			\footnotesize $D_1$  & \footnotesize $D_5$   &\footnotesize Ground Truth \\
		\end{tabular}
	\end{center}  
	\caption{Comparing the results of training with single dark level data and training with all dark level data. $D_1$ and $D_5$ represent the single-darkness and full-darkness data used by our method, respectively. Using data with separate darkness levels results in noticeable color distortion and artifacts.}
	\label{fig:1_or_5_compare}
\end{figure}

\noindent\textbf{Effectivenesses of the introducing different losses.}
To verify the effectiveness of each term of the total loss, we train our LoLiSRFlow on real dataset for $\times2$ super-resolution by using $\mathcal{L}_1$ loss, $\mathcal{L}_{nll}$ loss, and $\mathcal{L}_{total}$ loss, respectively. 
As shown in Table~\ref{tab:ablation}, using the total loss function compares to using $\mathcal{L}_1$ or $\mathcal{L}_{nll}$ alone obtains better qualitative and quantitative results, which also demonstrate the effectiveness of introducing $\mathcal{L}_{nll}$ loss as an additional constraint. As demonstrated in  Figure~\ref{fig:loss_compare}, we can easily see that our results maintain good artifact suppression properties.
\vspace{0.1cm}

\noindent\textbf{Adaptability of the model to different darkness levels.}
To verify that our model can adapt to data with different darkness levels, we train the model with data with darkness levels of $-3.0$ EV and all darkness levels separately. As shown in Table~\ref{tab:ablation}, training with only $-3.0$ EV darkness level data produces poor quantitative results. In Figure~\ref{fig:1_or_5_compare}, a network trained with only a single dark-level data produces color distortions and artifacts when processing other dark-level data. It demonstrates that the proposed method is adaptable and robust to different darkness levels.
\vspace{0.1cm}

\noindent\textbf{Effectivenesses of the CR map.}
To verify the effectiveness of the CR map, we perform ablation experiments on the RELLISUR dataset using our LoLiSRFlow. As shown in Table~\ref{tab:ablationCR}, better psnr and ssim metrics are achieved by using CR maps in the case of using different losses. This demonstrates the effectiveness of CR map. 

\begin{table}[htb]
	\centering
	\caption{Quantitative comparison of different losses and whether to use color ratio maps on RELLISUR dataset($\times2$).}
    \setlength{\tabcolsep}{3.5 mm}{
	\begin{tabular}{ccccc}
		\toprule
		\multicolumn{1}{c}{Using} &\multicolumn{2}{c}{Loss}&\multirow{2}{*}{PSNR $\uparrow$} & \multirow{2}{*}{SSIM $\uparrow$}\vspace{0.001in}\\		
		\cmidrule{2-3}
		CR Map & $L_1$ & $L_{nll}$ &  &  \\
		\midrule   
		\xmark & \xmark & \cmark & 19.07  & 0.649 \\
		\xmark & \cmark & \cmark & 22.96  & 0.768 \\
		\cmark & \xmark & \cmark & 22.94  & 0.747 \\
		\cmark & \cmark & \cmark & 23.41  & 0.783 \\
		\bottomrule
	\end{tabular}
}
\label{tab:ablationCR}  
\end{table}

\section{Conclusion}
In this paper, we propose a novel yet efficient flow architecture for simultaneously
solving the low-light enhancement and super-resolution tasks. We introduce a transformer-based cross-scale conditional encoder and a resolution- and illumination-invariant color ratio map as conditional priors with supervised constraints applied to the conditional encoding and image side. Furthermore, we propose DFSR-LLE, a dataset covering various low-light, low-resolution scenes using realistic noise models. Qualitative and quantitative experimental results show that our proposed method has a better performance compared to state-of-the-art techniques.

\begin{appendices}
%
%




\end{appendices}


\bibliography{egbib}
\end{document}